\documentclass[final,5p,times,twocolumn,authoryear]{elsarticle}

\usepackage{amssymb}
\usepackage{lipsum}
\usepackage{graphicx, times} 
\usepackage{natbib}
\usepackage{amssymb,amsmath}
\usepackage{threeparttable}
\usepackage{array}
\usepackage{multirow}
\usepackage{graphicx}
\usepackage{subcaption}
\usepackage[utf8]{inputenc}
\usepackage{amsthm}
\usepackage{lineno}

\DeclareUnicodeCharacter{2009}{\thinspace}

\journal{New Astronomy}

\begin{document}

\begin{frontmatter}

\title{Probing Dust and PAH Chemistry in Evolved Carbon-Rich Nebulae through Optical and Infrared Observations}

\author[label1]{Rahul Kumar Anand}
\author[label1]{Atul Kumar Singh}
\affiliation[label1]{organization={Department of Physics, Deen Dayal Upadhyaya Gorakhpur University},
            addressline={Civil Lines}, 
            city={Gorakhpur},
            postcode={273009}, 
            state={Uttar Pradesh},
            country={India}}
            
\author[label2]{Saurabh Sharma}   
\author[label2]{Brijesh Kumar}
 \affiliation[label2]{organization={Aryabhatta Research Institute of Observational Sciences},
            addressline={Manora Peak}, 
            city={Nainital},
            postcode={263001}, 
            state={Uttarakhand},
            country={India}}         

\author{Shantanu Rastogi\corref{cor1}\fnref{label1}}
\ead{shantanu_r@hotmail.com}

\begin{abstract}
This study presents optical and near-infrared photometric observations, alongside mid-infrared spectroscopic data from the ISO SWS instrument, to examine potential correlations between Aromatic Infrared Band (AIB) features and the optical properties of carbon-rich evolved stars. Identifying such correlations can provide valuable constraints on the evolutionary pathways of low- to intermediate-mass stars beyond the asymptotic giant branch (AGB) phase. Photometric measurements in the U, B, V, R, I, J, H, K, and L bands were obtained for five well-known carbon-rich objects at various post-AGB or planetary nebula (PN) stages: CRL 2688, PN M 2-43, NGC 7027, BD${+}$30${^\circ}$3639, and AFGL 2132. Our analysis reveals that all five objects exhibit prominent AIB features; however, their spectral profiles show notable variation. These differences are attributed to variations in the chemical composition and physical conditions of the surrounding circumstellar material. In particular, the 3.28$\mu$m polycyclic aromatic hydrocarbon (PAH) feature is detected in all objects except AFGL 2132, indicating a potentially distinct PAH population or environmental condition in its vicinity. Although these sources share broadly similar evolutionary stages, the observed diversity in AIB characteristics underscores the complexity and heterogeneity of their circumstellar environments.
\end{abstract}



\begin{keyword}
ISM: lines and bands \sep ISM: molecules \sep stars: AGB and post-AGB \sep planetary nebulae: general \sep techniques: spectroscopic and photometric
\end{keyword}

\end{frontmatter}


\section{Introduction}
\label{introduction}

Planetary nebulae (PNe) represent a transient yet crucial phase in the evolution of intermediate-mass stars (1–8 M$_\odot$). In this phase, the intense ultraviolet radiation from the hot central star ionizes the ejected circumstellar material, producing the rich emission-line spectra characteristic of PNe \citep{Kastner1996, Zuckerman1988, Hora1999}. Carbon-rich PNe (with $C/O > 1$) are of particular interest because they synthesize and expel copious amounts of carbonaceous dust and complex organic molecules during the late AGB and post-AGB stages. Observationally, many PNe exhibit pronounced departures from spherical symmetry, often taking on bipolar or multi-polar morphologies, which likely reflect shaping processes (e.g., binary central stars or magnetic fields) during the nebular ejection \citep{Huggins1989, Huggins1996}. Such asymmetric structures are especially common in post-AGB nebulae and young PNe, and they can lead to an uneven distribution of dust and gas around the star. This, in turn, affects both the SED and the extinction across the nebula, underscoring the importance of studying PNe across multiple wavelengths to capture both their ionized and dusty components.

Carbon-rich evolved nebulae are also distinctive due to the presence of mid-infrared emission features known as the Aromatic Infrared Bands (AIBs). Space infrared observatories (IRAS, ISO, Spitzer) have revealed ubiquitous emission features at wavelengths $\sim$3.3, 6.2, 7.7, 8.6, 11.2, and 12.7 $\mu$m in the spectra of PNe. These features are attributed to vibrational modes of PAH molecules excited by the UV radiation from the central star\citep{Leger1984, Allamandola1989}. The emission bands (3.3$\mu$m for aromatic C–H stretch, 6.2/7.7/8.6 $\mu$m for aromatic C–C and C-H in-plane modes, and 11.2$\mu$m for aromatic out-of-plane C–H bending) serve as fingerprints of PAHs in these environments \citep{Tielens2008, Rastogi2013}. Over the past decades, infrared spectral surveys have shown that profiles and peak positions of these PAH features can vary from object to object. In fact, PAH emission spectra have been empirically classified into different types (e.g., “Class A,” “B,” or “C”) based on subtle shifts in band positions and shapes \citep{Peeters2002, Diedenhoven2004}. This classification correlates with the nature of the source: for instance, PAH spectra in nebulae around hot, UV-bright central stars have more pronounced 6–8 $\mu$m complexes than those in cooler or less excited environments. Recent high-resolution observations have reinforced the ubiquity and diagnostic value of PAH emission in PNe. For example, JWST imaging of the Ring Nebula (NGC~6720) unveiled a thin ring of PAH emission encircling the ionized gas shell \citep{Wesson2024}, directly illustrating how PAH-rich material is distributed in these nebular environments. As the star evolves into the PN phase, the dust remains as a cold (tens to hundreds of K) component intermixed with the molecular gas, and it re-radiates absorbed starlight in the infrared. A significant fraction of a PN’s luminosity is often emitted in the infrared continuum by these dust grains \citep{Kwok2022}. 

The presence of a dust envelope around the central star of PNe has two major observable effects: (1) it causes extinction of the optical light from the nebula and central star, often measurable as a reddening of the nebular continuum and line ratios; and (2) it contributes excess emission at infrared wavelengths. In the present study, both effects are important. By measuring the nebular colors in optical and near-infrared bands, one can infer the line-of-sight extinction $A_V$ for each object. Despite considerable progress, many aspects of the dust and PAH evolution in the late stages of stellar evolution remain uncertain. The transition from the AGB to PN phase is rapid (a few thousand years or less), and during this brief period, the circumstellar environment undergoes dramatic changes: the central star’s temperature rises, the ionization front expands, and shocks and fast winds can overtake slower AGB ejecta. How these processes affect the survival and chemistry of dust grains and PAH molecules is an open question. \citep{Frenklach1989, Cherchneff1992, Tielens2008}. If systematic differences in PAH emission or dust properties are identified and correlated with the evolutionary status, it can illuminate the chemical pathways at play.

The present work probes the dust and PAH chemistry in a sample of five carbon-rich evolved nebulae by combining optical and infrared observations. The targets – AFGL~2132, CRL~2688, PN~M~2-43, NGC~7027, and BD $+$30$^\circ$ 3639 – span the evolutionary sequence from the post-AGB stage to young planetary nebulae. All are known to exhibit strong AIB emissions and are enshrouded by significant dust. Briefly, CRL~2688 is largely seen in reflected light and molecular emission \citep{Young1992, Latter1993, Cox1997}. It bipolar post-AGB object believed to be in a short-lived transition phase between the AGB and PN stages. PN M~2-43, NGC~7027, and BD $+$30$^\circ$ 3639 are young PNe with compact, high-density nebulae; NGC~7027 and BD $+$30$^\circ$ 3639 in particular host hydrogen-deficient [WC] type central stars and are well-studied prototypes for PAH-rich nebulae. AFGL~2132 is an emission-line star with a strikingly symmetric dusty nebula; its exact evolutionary nature is unclear, either a very young PN or an unusual binary outflow. It has extremely heavy obscuration along the line of sight \citep{Contreras2017}. The study aims to identify common trends and notable differences in their dust and PAH characteristics of these objects. Broadband photometric observations are obtained in the optical ($U, B, V, R, I$) and near-infrared ($J, H, K$ and narrow-band L and PAH 3.3~$\mu$m ) filters for these objects, using the 1.3-m and 3.6-m telescopes at Devasthal (ARIES, India). These data provide calibrated colors and magnitudes, from which extinction values are derived and color excess is investigated. Archival mid-infrared spectra (ISO/SWS) for each nebula, which cover the key PAH emission features in the 2.4–15~$\mu$m  range, are also used. By combining the optical extinction analysis with the PAH spectral profiles, we seek to correlate the optical properties (e.g., derived $E(B-V)$, nebular morphology) with the PAH emission characteristics (e.g., the relative strength of the 3.3 vs 11.2~$\mu$m  bands). Any discovered correlations (or lack thereof) will inform how the dust and PAH chemistry evolves as the star leaves the AGB.

\section{Observations and Data Analysis}
\label{sect:Obs}
\subsection{Optical Photometry}
The observations in Johnson U, B, V, and Cousins R, I filters of five C-rich objects were obtained by using 1.3-m Devasthal Fast Optical Telescope (DFOT) \citep{Sagar2011} at Devasthal, India during the night of 15-Oct-2017. The telescope is situated at Devasthal (Longitude 79$^\circ$ 41$^{\prime}$ 04$^{\prime\prime}$ E, Latitude 29$^\circ$21$^{\prime}$40$^{\prime\prime}$N, Altitude: 2450 m), India, and operated by the Aryabhatta Research Institute of Observational Sciences (ARIES), Nainital. During the observations, the telescope was equipped with a 2k$\times$2k CCD camera having a field of view of $\sim$ 18$^{\prime}$ $\times$ 18$^{\prime}$. The CCD has gain 2 e\textsuperscript{-}/ADU and read noise is 7 e\textsuperscript{-}. The basic data reduction (image cleaning, photometry, and astrometry) was done using the standard IRAF procedure \citet{2020MNRAS.498.2309S,2017MNRAS.467.2943S}. The SA98 standard star field \citep{Landolt1992}  was observed during the same night for calibration of the target photometry. The observation log for photometric data is given in Table~\ref{tbl-obs_log}.

\begin{table*}
\centering
\scriptsize
\begin{threeparttable}
\renewcommand{\arraystretch}{1.2}
\caption{Combined log of optical and near-IR observations of the five objects. Optical bands (U, B, V, R, I) were observed using the 1.3-m DFOT, and near-IR bands(J, H, K, nbl, PAH) using the 3.6-m DOT.}
\label{tbl-obs_log}

\begin{tabular}{c}
\hline
\textbf{AFGL 2132} (Emission-line star) \\
RA: 18:21:16.06 \quad DEC: $-$13:01:25.6 \\
\hline
\begin{tabular}{cccccccccccc}
Filter &U& B & V & R & I & J & H & K & PAH & nbL \\
Exposure &  &1$\times$120$\times$3 & 1$\times$20$\times$3 & 1$\times$5$\times$3 & 1$\times$5$\times$3 & 
4$\times$5$\times$5 & 4$\times$5$\times$5 & 3$\times$10$\times$0.05 & 4$\times$300$\times$0.05 & 4$\times$300$\times$0.05 \\
\end{tabular} \\
\hline
\textbf{CRL 2688} (Post-AGB star) \\
RA: 21:02:18.27 \quad DEC: +36:41:37.0 \\
\hline
\begin{tabular}{cccccccccccc}
Filter&U & B & V & R & I & J & H & K & PAH & nbL \\
Exposure &1$\times$60$\times$3 & 1$\times$25$\times$3 & 1$\times$10$\times$3 & 1$\times$5$\times$3 & 1$\times$5$\times$3 & 
3$\times$3$\times$10 & 3$\times$3$\times$10 & 3$\times$6$\times$5 & 3$\times$300$\times$0.05 & 3$\times$300$\times$0.05 \\
\end{tabular} \\
\hline
\textbf{PN M 2-43} (Planetary Nebula) \\
RA: 18:26:40.05 \quad DEC: $-$02:42:57.3 \\
\hline
\begin{tabular}{cccccccccccc}
Filter&U &B & V & R & I & J & H & K & PAH & nbL \\
Exposure& $ ~ ~ ~ ~ ~ ~ ~ $&1$\times$180$\times$3 & 1$\times$40$\times$3 & 1$\times$15$\times$3 & 1$\times$15$\times$3 & 
3$\times$5$\times$40 & 3$\times$10$\times$10 & $ ~ ~ ~ ~ ~ ~ ~ ~   $ &3$\times$300$\times$0.05 & 3$\times$300$\times$0.05 \\
\end{tabular} \\
\hline

\textbf{NGC 7027} (Planetary Nebula) \\
RA: 21:07:01.53 \quad DEC: +42:14:11.5 \\
\hline
\begin{tabular}{cccccccccccc}
Filter&U & B & V & R & I & J & H & K & PAH & nbL \\
Exposure &1$\times$30$\times$3 & 1$\times$20$\times$3 & 1$\times$10$\times$3 & 1$\times$5$\times$3 & 1$\times$5$\times$3 & 
3$\times$4$\times$20 & 3$\times$4$\times$20 & 3$\times$7$\times$12 & 3$\times$300$\times$0.05 & 3$\times$300$\times$0.05 \\
\end{tabular} \\
\hline

\textbf{BD+30$^\circ$3639} (Planetary Nebula) \\
RA: 19:34:45.30 \quad DEC: +30:30:59.2 \\
\hline
\begin{tabular}{cccccccccccc}
Filter&U & B & V & R & I & J & H & K & PAH & nbL \\
Exposure &1$\times$30$\times$3 & 1$\times$20$\times$3 & 1$\times$10$\times$3 & 1$\times$5$\times$3 & 1$\times$5$\times$3 & 
3$\times$3$\times$8 & 3$\times$3$\times$8 & 3$\times$5$\times$5 & 3$\times$300$\times$0.05 & 3$\times$300$\times$0.05 \\
\end{tabular} \\
\hline
\end{tabular}
\begin{tablenotes}\footnotesize

\item Note: Exposure time is given as (Dither Positions × Number of Frames × Seconds).
\item ~ ~ ~ ~ ~ ~RA and DEC are in the format (hh:mm:ss) and (dd:mm:ss), respectively, for epoch J2000.
\item ~ ~ ~ ~ ~ ~Blank space denoted object is not observed in that band.
\end{tablenotes}
\end{threeparttable}
\end{table*}

The instrumental magnitudes were transformed to the standard photometric system using the following calibration equations, as described by \citet{Stetson1992}.
\begin{align}
    U - B = 0.83\times(u-b)\textsubscript{0} - 1.26 \\
    B - V = 1.27\times(b-v)\textsubscript{0} - 0.82 \\
    V - R = 0.95\times(v-r)\textsubscript{0} - 0.25 \\
    V - I = 0.93\times(v-i)\textsubscript{0} + 0.27 \\
    V - v\textsubscript{0} = -0.09\times(V-R) - 2.23
\end{align}

Here, $U$, $B$, $V$, $R$, and $I$ represent the standard magnitudes, and $u$, $b$, $v$, $r$, and $i$ represent the instrumental magnitudes which are normalized for the corresponding exposure time and corrected for extinction in that band ($\bar{k}_\lambda$). The mean extinction values for the Devasthal site as reported by \citet{Brijesh2000},  $\bar{k}_u = 0.49 \pm 0.09$, $\bar{k}_b = 0.32 \pm 0.06$, $\bar{k}_v = 0.21 \pm 0.05$, $\bar{k}_r = 0.13 \pm 0.04$, and $\bar{k}_i = 0.08 \pm 0.04$ for the U, B, V, R, and I bands, respectively.
The resulting calibrated magnitudes of the objects are listed in Table~\ref{tbl-photometry}.

\subsection{NIR Photometry}
The near infrared imaging of these objects in J, H, K, and narrow band L \& PAH bands were obtained with the  TIFR Near Infrared Imaging Camera-II (TIRCAM2) on the 3.6-m Devasthal Optical Telescope (DOT), ARIES, Nainital, during the nights of 10-May-2018 and 13-Oct-2018. The TIRCAM2 is a closed-cycle cryocooled near-IR imaging camera equipped with a larger Raytheon 512 x 512 pixels InSb Aladdin III Quadrant focal plane array (FPA) \citep{Naik2012, Baug2018}, and it was attached to the main axial port of the Cassegrain focus of the 3.6-m DOT during the observations. 

A standard observing and data reduction strategy was adopted for the near-IR observations, following the procedures described in \citet{2020MNRAS.498.2309S}.

Standard stars \citep{Hunt1998} were also observed in J, H, and K bands during each night for calibration. The observation log for near-IR data is shown in Table~\ref{tbl-obs_log}. The instrumental magnitudes of the target sources were calibrated using the observed standard sources \citep{Hunt1998}. The calibrated magnitudes of the target sources are presented in Table~\ref{tbl-photometry}.

\begin{table*}
\begin{center}
\scriptsize
\begin{threeparttable}
\renewcommand{\arraystretch}{1.1}
\caption{UBVRI optical and JHK near-infrared photometric magnitudes of five carbon-rich objects. The table includes both observed JHK magnitudes and those retrieved from the 2MASS catalog. The final column indicates the detection of the 3.28~$\mu$m PAH emission feature based on our observed data.}\label{tbl-photometry}

\begin{tabular}{|m{1.6cm}|
              m{0.8cm}m{0.8cm}m{0.8cm}m{0.8cm}m{0.8cm}|
              m{0.8cm}m{0.8cm}m{0.8cm}|
              m{0.5cm}m{0.5cm}m{0.5cm}|
              m{1.0cm}|}
\hline
\multirow{2}{*}{Object} 
    & \multicolumn{5}{c|}{UBVRI (Observed)} 
    & \multicolumn{3}{c|}{JHK (Observed)} 
    & \multicolumn{3}{c|}{JHK (2MASS)} 
    & Observed PAH at \\
    \cline{2-12}
    & U & B & V & R & I & J & H & K & J & H & K & 3.28$\mu$m \\
    & Mag & Mag & Mag & Mag & Mag & Mag & Mag & Mag & Mag & Mag & Mag & \\
\hline
AFGL 2132 
    &  & 15.930 & 14.160 & 12.690 & 11.244 
    & 10.226 & 8.924 & 9.257 
    & 8.96 & 7.396 & 5.704 & No \\
    &     & $\pm$0.011 & $\pm$0.007 & $\pm$0.004 & $\pm$0.002 
    & $\pm$0.005 & $\pm$0.013 & $\pm$0.009 
    &       &       &       &     \\
\hline
CRL 2688 
    & 13.022 & 12.272 & 11.298 & 10.682 & 10.034 
    & 10.794 & 10.247 & 8.580 
    & 9.865 & 9.352 & 8.839 & Yes \\
    & $\pm$0.008 & $\pm$0.002 & $\pm$0.002 & $\pm$0.002 & $\pm$0.001 
    & $\pm$0.002 & $\pm$0.005 & $\pm$0.005 
    &       &       &       &     \\
\hline
PN M 2-43 
    &  & 16.114 & 15.194 & 13.640 & 13.017 
    & 10.094 & 9.949 &  
    & 11.343 & 10.532 & 8.91 & Yes \\
    &     & $\pm$0.007 & $\pm$0.006 & $\pm$0.003 & $\pm$0.003 
    & $\pm$0.004 & $\pm$0.005 &  
    &       &       &       &     \\
\hline
NGC 7027 
    & 10.354 & 9.294 & 8.523 & 8.594 & 8.672 
    & 6.860 & 6.767 & 7.010\tnote{*} 
    & 9.736 & 8.260 & 7.305 & Yes \\
    & $\pm$0.003 & $\pm$0.001 & $\pm$0.001 & $\pm$0.001 & $\pm$0.001 
    & $\pm$0.003 & $\pm$0.004 & $\pm$0.010 
    &       &       &       &     \\
\hline
BD $+$30$^\circ$~3639 
    & 8.970 & 9.581 & 9.862 & 8.329 & 9.078 
    & 9.715 & 9.443 & 8.208 
    & 9.306 & 9.231 & 8.108 & Yes \\
    & $\pm$0.001 & $\pm$0.001 & $\pm$0.001 & $\pm$0.001 & $\pm$0.001 
    & $\pm$0.005 & $\pm$0.012 & $\pm$0.009 
    &       &       &       &     \\
\hline
\end{tabular}
\begin{tablenotes}\footnotesize
\item[*] K-band magnitude of NGC 7027 is taken from \citet{Taranova2007}.
\end{tablenotes}
\end{threeparttable}
\end{center}
\end{table*}

\subsection{Archived Data}
\subsubsection{2MASS Data}
Near-IR magnitude data in J, H, and K$\textsubscript{s}$ from the Two Micron All-Sky Survey (2MASS) archive \citep{Skrutskie2006} are used. The archived data have a signal-to-noise ratio (S/N) greater than 10 with limiting magnitudes of 15.8, 15.1, and 14.3 mag in J, H, and K$\textsubscript{s}$, respectively. The J, H, K$\textsubscript{s}$ magnitudes from 2MASS are also listed in Table~\ref{tbl-photometry}.
 
\subsubsection{ISO Data}
The mid-infrared spectrum data were obtained from the Infrared Space Observatory (ISO) archive; all mid-IR data were acquired by the Short Wavelength Spectrometer (SWS) \citep{Graauw1996} covering 2.4 - 45.4 $\mu$m regions. 
The background continuum is subtracted from each spectrum, using a spline curve, to obtain a baseline and relative intensity of absorption features. The background-subtracted mid-IR spectra for all five objects are shown in Figure 2. Flux ratios between 6.2 and 11.2 $\mu$m are shown in Table~\ref{tbl-aromatic_bands}.

\begin{table*}
\begin{center}
\scriptsize
\begin{threeparttable}
\renewcommand{\arraystretch}{1.5}
\caption{Comparison of observed and 2MASS-derived logarithmic flux densities (in $erg~s^{-1}~cm^{-3}$) for J, H, and K bands for the five target objects. The logarithm of the effective wavelength (in $\mu$m) for each band is also listed.}\label{tbl-flux-revised}
\begin{tabular}{|p{0.6cm}|p{0.7cm}|p{0.7cm}|p{0.7cm}|p{0.7cm}|p{0.7cm}|p{0.7cm}|p{0.7cm}|p{0.7cm}|p{0.7cm}|p{0.7cm}|p{0.7cm}|}
\hline
\multirow{3}{*}{Band} & \multirow{3}{*}{Log$\lambda$} & \multicolumn{10}{c|}{Log F (erg s$^{-1}$ cm$^{-3}$)} \\
\cline{3-12}
 & & \multicolumn{2}{c|}{AFGL 2132} & \multicolumn{2}{c|}{CRL 2688} & \multicolumn{2}{c|}{PN M 2-43} & \multicolumn{2}{c|}{NGC 7027} & \multicolumn{2}{c|}{BD$+$30$^\circ$~3639} \\
\cline{3-4} \cline{5-6} \cline{7-8} \cline{9-10} \cline{11-12}
 & ($\mu$m) & Obs & 2MASS & Obs & 2MASS & Obs & 2MASS & Obs & 2MASS & Obs & 2MASS \\
\hline
J & -0.09 & -5.59 & -5.09 & -5.82 & -5.78 & -5.54 & -6.04 & -4.25 & -5.41 & -5.39 & -5.23 \\
H & -0.22 & -5.52 & -4.91 & -6.05 & -6.07 & -5.93 & -6.17 & -4.66 & -5.31 & -5.73 & -5.65 \\
K & -0.33 & -6.07 & -4.65 & -5.80 & -6.42 &       & -5.93 & -5.17 & -5.34 & -5.65 & -5.93 \\
\hline
\end{tabular}
\end{threeparttable}
\end{center}
\end{table*}

\subsection{Extinction determination}

To estimate the interstellar extinction for each object, we adopted observed broadband colors (from UBVRI and JHK photometry) and compared them with expected intrinsic colors based on the object’s spectral type. The color excess $E(X-Y)$ is calculated as the difference between the observed color $(X-Y)_{\text{obs}}$ and the intrinsic color $(X-Y)_0$. The primary optical color excess $E(B-V)$ is used to derive the visual extinction $A_V$ via $A_V = R_V \times E(B-V)$. The standard extinction law with $R_V = 3.1$ \citep{Cardelli1989} is adopted, which is appropriate for average interstellar dust. For near infrared bands, the extinction ratios from \citet{Rieke1985} are used to compute $A_J$, $A_H$, and $A_K$; specifically, $A_J/A_V \approx 0.282$, $A_H/A_V \approx 0.175$, and $A_K/A_V \approx 0.112$. These ratios can be used to calculate band-specific extinctions once $A_V$ is known (see Table~\ref{tab:combined_colors}).

\section{Results}

\subsection{AFGL 2132}
\begin{figure*}[h]
     \centering
     \begin{subfigure}[b]{0.24\textwidth}
     \includegraphics[width=\textwidth]{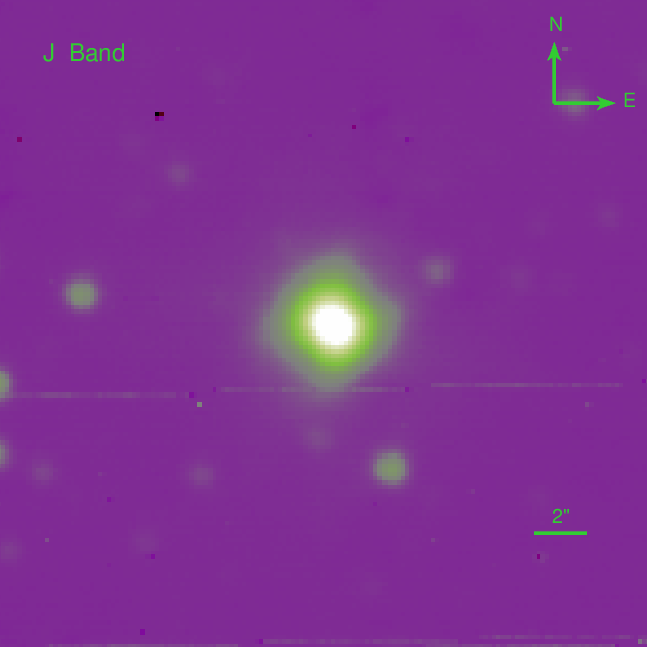}
     \end{subfigure}
     \hfill
     \begin{subfigure}[b]{0.24\textwidth}
         \includegraphics[width=\textwidth]{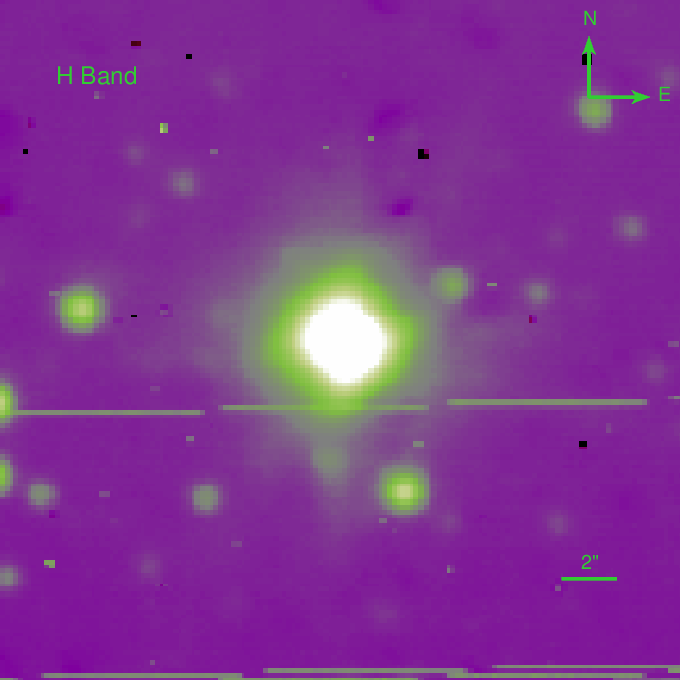}    
     \end{subfigure}
     \hfill
     \begin{subfigure}[b]{0.24\textwidth}
         \includegraphics[width=\textwidth]{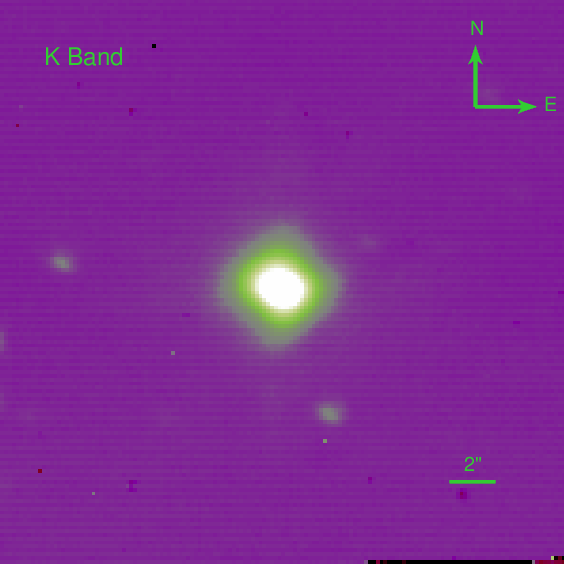}    
     \end{subfigure}
     \hfill
     \begin{subfigure}[b]{0.24\textwidth}
         \includegraphics[width=\textwidth]{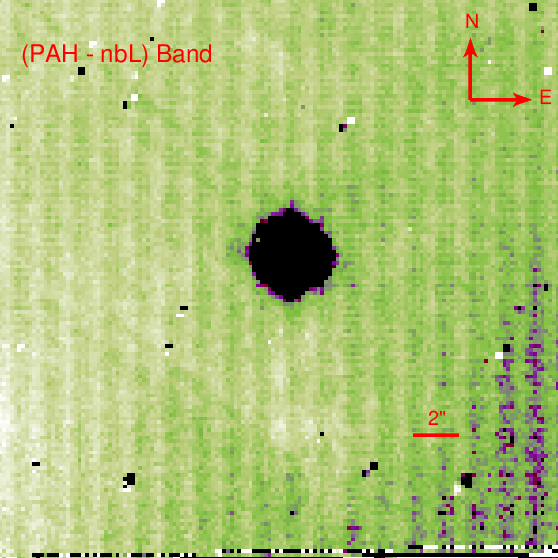}    
     \end{subfigure}     
        \includegraphics[width=0.48\textwidth]{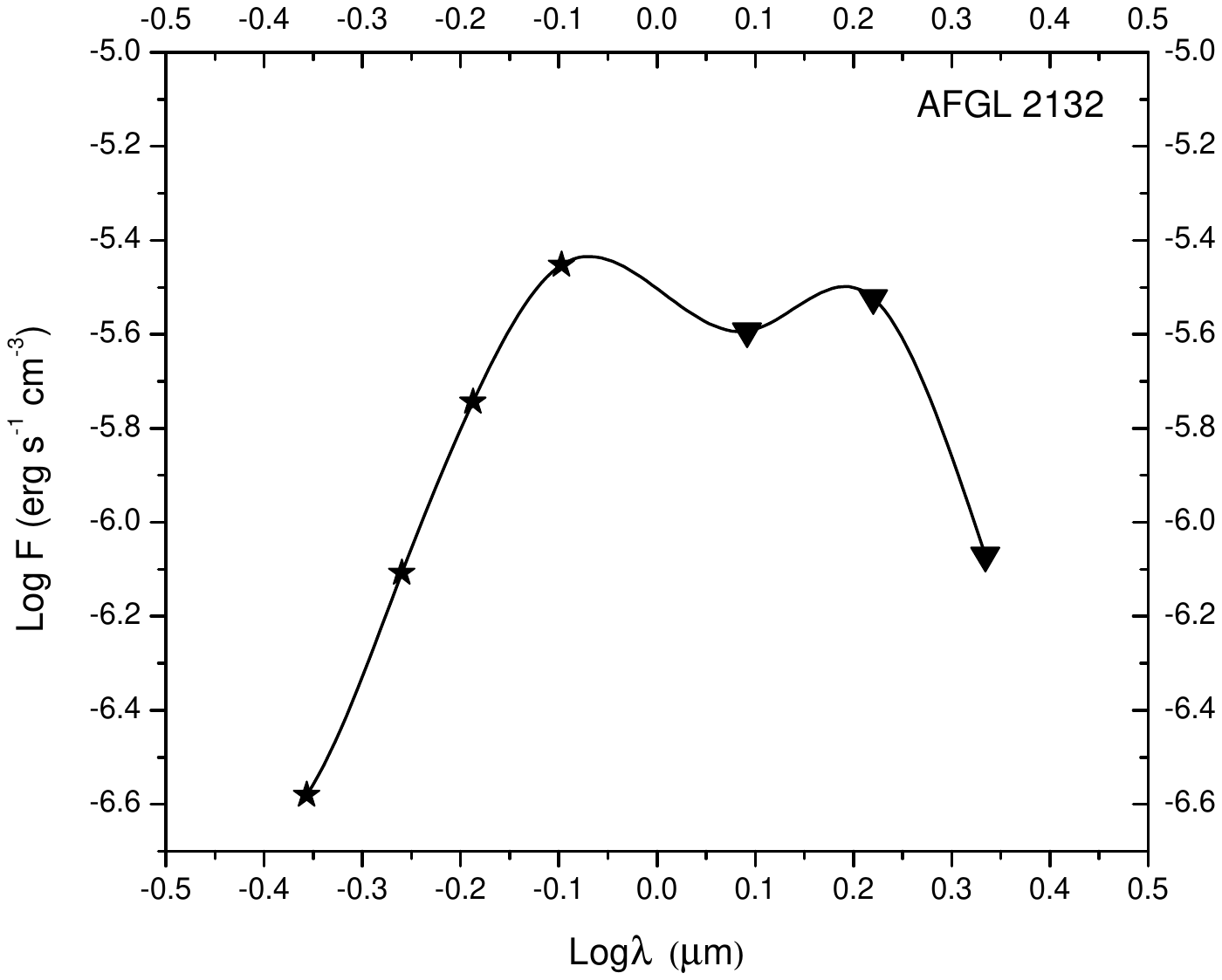}
     \includegraphics[width=0.48\textwidth]{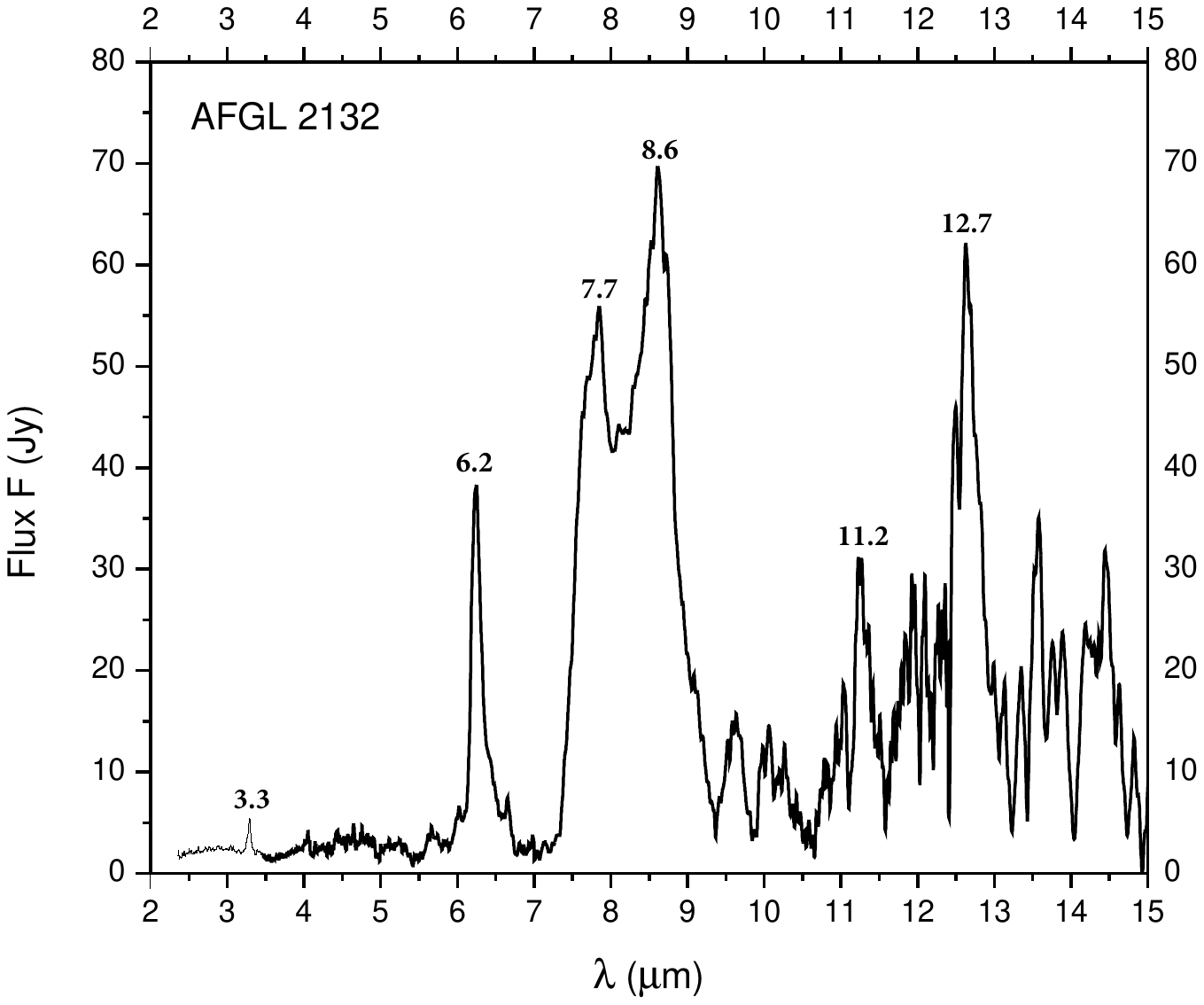}
     \caption{ \textbf{AFGL 2132} Upper panels: Near-IR imaging in different bands. Lower panels: BVRIJHK spectral energy distribution (left), BVRI are denoted with $\star$ while JHK are denoted with $\blacktriangledown$ and background-subtracted mid-IR emission features from ISO SWS (right), major aromatic infrared bands are indicated in the spectra.
      }
\label{fig-afgl2132}
\end{figure*}

AFGL 2132, also designated as MWC 922 and commonly known as the ``Red Square Nebula," is classified as an emission-line B[e] star enshrouded in a thick circumstellar dust environment \citep{Allen1976}. High-resolution near-infrared imaging in the $JHK$ bands reveals a remarkably symmetrical square morphology with dimensions approximately 5.07$\times$5.07~arcsec (Figure~\ref{fig-afgl2132}). The central bright region, clearly visible in these infrared images, measures approximately 1.86$\times$1.86~arcsec. Photometric measurements give a visual magnitude of $V=14.16\pm0.007$, with significantly reddened colors of $(B-V)=+1.77$ and an unusually negative infrared color of $(H-K)=-0.333$. 

The nebula exhibits pronounced continuum emission at 3.59~$\mu$m (nbL-band), whereas the AIB at 3.28~$\mu$m appears negligible in the PAH--nbL difference image (Figure~\ref{fig-afgl2132}). This absence of strong 3.28~$\mu$m emission is typical in environments with intense ultraviolet radiation fields, such as those surrounding early-type emission-line stars, where (PAHs) become predominantly ionized and the C--H stretching features weaken significantly \citep{Pathak2008}.

The SED constructed from BVRIJHK photometry resembles a blackbody profile, indicative of thermal emission from circumstellar dust. The peak emission corresponds to dust temperatures in the range 2000--4000~K, with substantial extinction particularly noticeable in the $J$ band (Figure~\ref{fig-afgl2132}).

Mid-infrared spectroscopy from ISO SWS reveals prominent emission features at 3.3, 6.2, 7.7, 8.6, and 11.2~$\mu$m (Figure~\ref{fig-afgl2132}, Table~\ref{tbl-aromatic_bands}). Following the classification scheme of \citet{Peeters2002} and \citet{Diedenhoven2004}, the 7.7~$\mu$m feature in AFGL 2132 falls into Class~A, while other bands match type B1. The 8.6~$\mu$m emission is notably strong and remains unclassified \citep{Peeters2002}.

Spectroscopic classification suggests a central star of spectral type B3--B6, intrinsically characterized by a blue continuum with an expected intrinsic $(B-V)_0\approx-0.2$ \citep{Ducati2001}. The significant discrepancy between the observed and intrinsic colors implies strong extinction. Using the observed color indices and standard extinction laws, we quantify this extinction.

\begin{table*}[h]
\scriptsize
\centering
\caption{Observed and intrinsic colors for the five objects, along with derived color excesses.}
\label{tab:combined_colors}
\begin{tabular}{|cccccccc|}
\hline
\multirow{2}{*}{Object} & \multicolumn{3}{c}{$(B-V)$} & \multicolumn{3}{c}{$(J-K)$} & References  \\
\cline{2-4} \cline{5-7}
 & Observed & Intrinsic & $E(B-V)$ & Observed & Intrinsic & $E(J-K)$ &  \\
\hline
AFGL~2132 & $+1.77$ & $-0.20$ & $1.97$ mag & $+0.97$ & $\sim0.0$ & $0.97$ mag & \cite{Ducati2001} \\
CRL~2688  & $+0.97$ & $+0.48$ & $0.49$ mag & $+2.21$ & $\sim0.0$ & $2.21$ mag & \cite{Johnson1966} \\
PN~M~2-43 & $+0.92$ & $-0.30$ & $1.22$ mag & $+2.43$ & $\sim0.0$ & $2.43$ mag & \cite{Vanderhucht2001}  \\
NGC~7027  & $+0.77$ & $-0.30$ & $1.07$ mag & $-0.15$ & $\sim0.0$ & N/A & \cite{Latter2000} \\
BD~$+$30$^\circ$~3639 & $-0.28$ & $-0.30$ & $\sim0.0$ & $+1.51$ & $\sim0.0$ & $1.51$ mag & \cite{Bernard-Salas2003} \\
\hline
\end{tabular}
\end{table*}

\begin{table*}[ht]
\centering
\scriptsize
\caption{Optical and NIR extinction estimates for the five objects. \cite{Cardelli1989} standard extinction law and extinction ratio from \cite{Rieke1985} are used.}
\label{tab:av_extinction_summary}
\begin{tabular}{|lccccc|}
\hline
\textbf{Object} & $E(B-V)$ & $A_V$ (mag) & $A_J$ (mag) & $A_H$ (mag) & $A_K$ (mag) \\
\hline
AFGL~2132 & 2.0 & 6.10 & 1.72 & 1.06 & 0.69  \\
CRL~2688 & 0.49 & 1.52 & 0.43 & 0.27 & 0.17  \\
PN~M~2-43 & 1.23 & 3.80 & 1.07 & 0.66 & 0.43  \\
NGC~7027 & 1.06 & 3.30 & 0.93 & 0.58 & 0.37  \\
BD+30$^\circ$~3639 & 0.34 & 1.05 & 0.30 & 0.18 & 0.12  \\
\hline
\end{tabular}
\end{table*}

Our derived extinction ($A_V\approx6.1$~mag) aligns closely with previous studies, which report similar levels of significant reddening toward AFGL 2132. For example, \citet{Voors1999} estimated $E(B-V)\approx2.0$ (corresponding to $A_V\approx6.2$~mag assuming $R_V=3.1$). It is important to note that our extinction values do not differentiate explicitly between interstellar and circumstellar dust components. Recent radio observations of free-free continuum and recombination lines suggest the presence of a massive compact binary companion, which could produce rapid winds, enhanced gas dynamics, and significant PAH ionization \citep{Contreras2019}, potentially contributing to the complex dust environment and the observed extinction.

\begin{table*}[h]
\scriptsize
\begin{center}
\caption{Mid-infrared PAH emission feature fluxes (in Jy) for the five target nebulae, derived from archival ISO spectra. The listed fluxes correspond to the major aromatic infrared bands at 3.3, 6.2, 7.7, 8.6, and 11.2~$\mu$m. The final column gives the 11.2~$\mu$m / 6.2~$\mu$m flux ratio, which serves as a diagnostic of PAH size and ionization state.}\label{tbl-aromatic_bands}

\begin{tabular}{|c|c|c|c|c|c|c|}
\hline
{Object}	&Flux (3.3 $\mu$m)	&Flux (6.2$\mu$m)	&Flux (7.7$\mu$m)	&Flux (8.6$\mu$m)	&Flux (11.2 $\mu$m ) & Flux ratio\\
						&	(Jy)			&(Jy)            	 	&(Jy)			 		&(Jy)         	&(Jy)  & (11.2$\mu$m / 6.2$\mu$m)  \\
\hline
AFGL 2132						&5.440	 				&38.296				&55.916				&69.729					&31.189 & 0.814\\
CRL 2688						&1.177					&10.480			     &NA			     &   NA						&29.689 & 2.833\\
PN M 2-43						&1.852					&8.919				&16.384				&7.328						&9.664 & 1.084\\
NGC 7027						&30.674		    		&73.779				&272.219				&65.232					&291.117 & 3.946\\
BD +30 3639			  		    &7.862					&36.978				&34.719				&23.018					&48.678 & 1.316\\
\hline
\end{tabular}
\end{center}
\end{table*}

\subsection{CRL 2688}
CRL~2688, commonly known as the Egg Nebula, is a bipolar protoplanetary nebula exhibiting a remarkable morphology, including twin searchlight beams and concentric arcs surrounding the central star \citep{Trung2020, Sahai1998}. Its central star is classified as spectral type F5Ia \citep{Crampton1975}, likely a post-AGB supergiant star, obscured from direct view by a dense equatorial dust torus. Photometric measurements give a visual magnitude $V=11.23$, with observed color indices $(B-V)=+0.97$ and $(U-B)=+0.75$, indicating significant internal reddening (see Table~\ref{tab:combined_colors}). 

\begin{figure*}[h]
     \centering
     \begin{subfigure}[b]{0.24\textwidth}
         \centering
         \includegraphics[width=\textwidth]{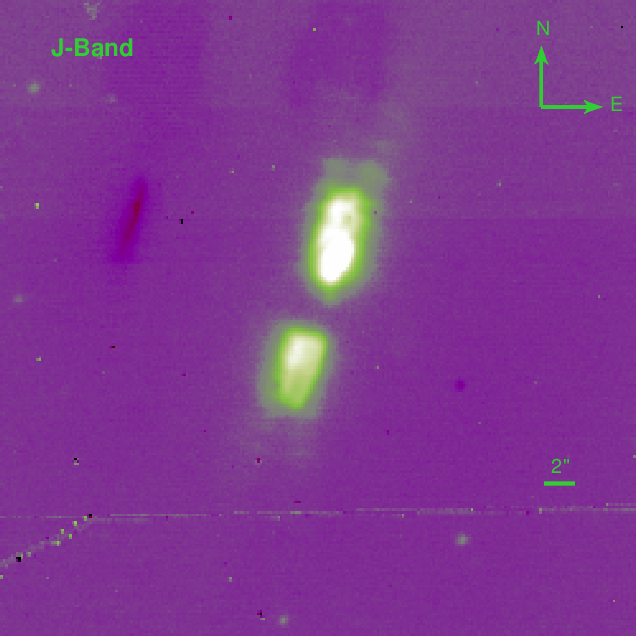}
        
     \end{subfigure}
     \hfill
     \begin{subfigure}[b]{0.24\textwidth}
         \centering
         \includegraphics[width=\textwidth]{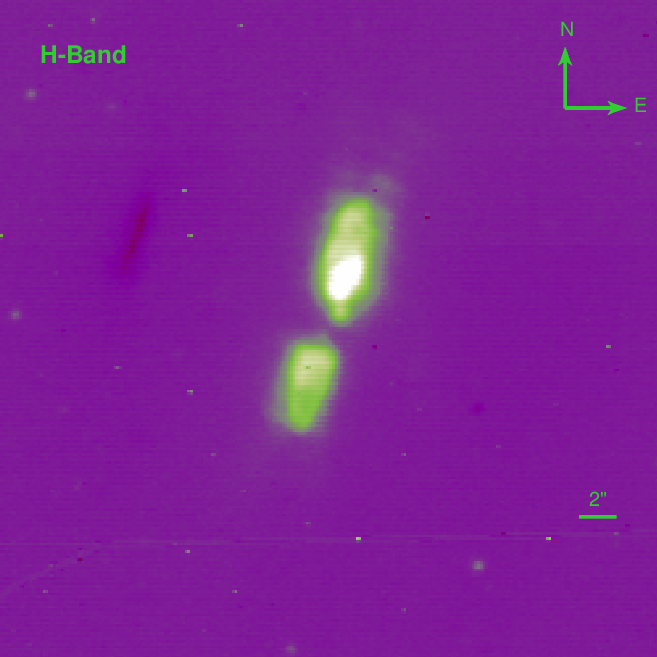}
         
     \end{subfigure}
     \hfill
     \begin{subfigure}[b]{0.24\textwidth}
         \centering
         \includegraphics[width=\textwidth]{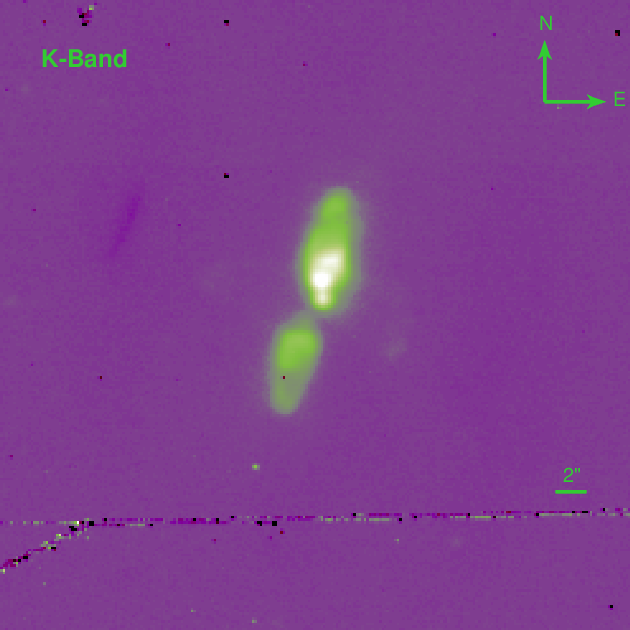}
        
     \end{subfigure}
     \hfill
     \begin{subfigure}[b]{0.24\textwidth}
         \centering
         \includegraphics[width=\textwidth]{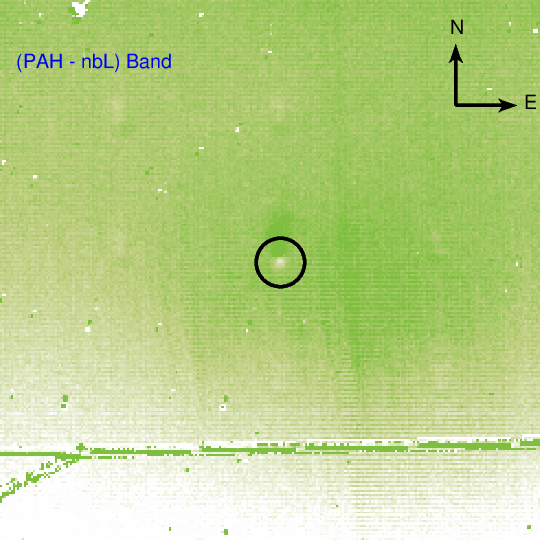}
    
     \end{subfigure}    
    \includegraphics[width=0.48\textwidth]{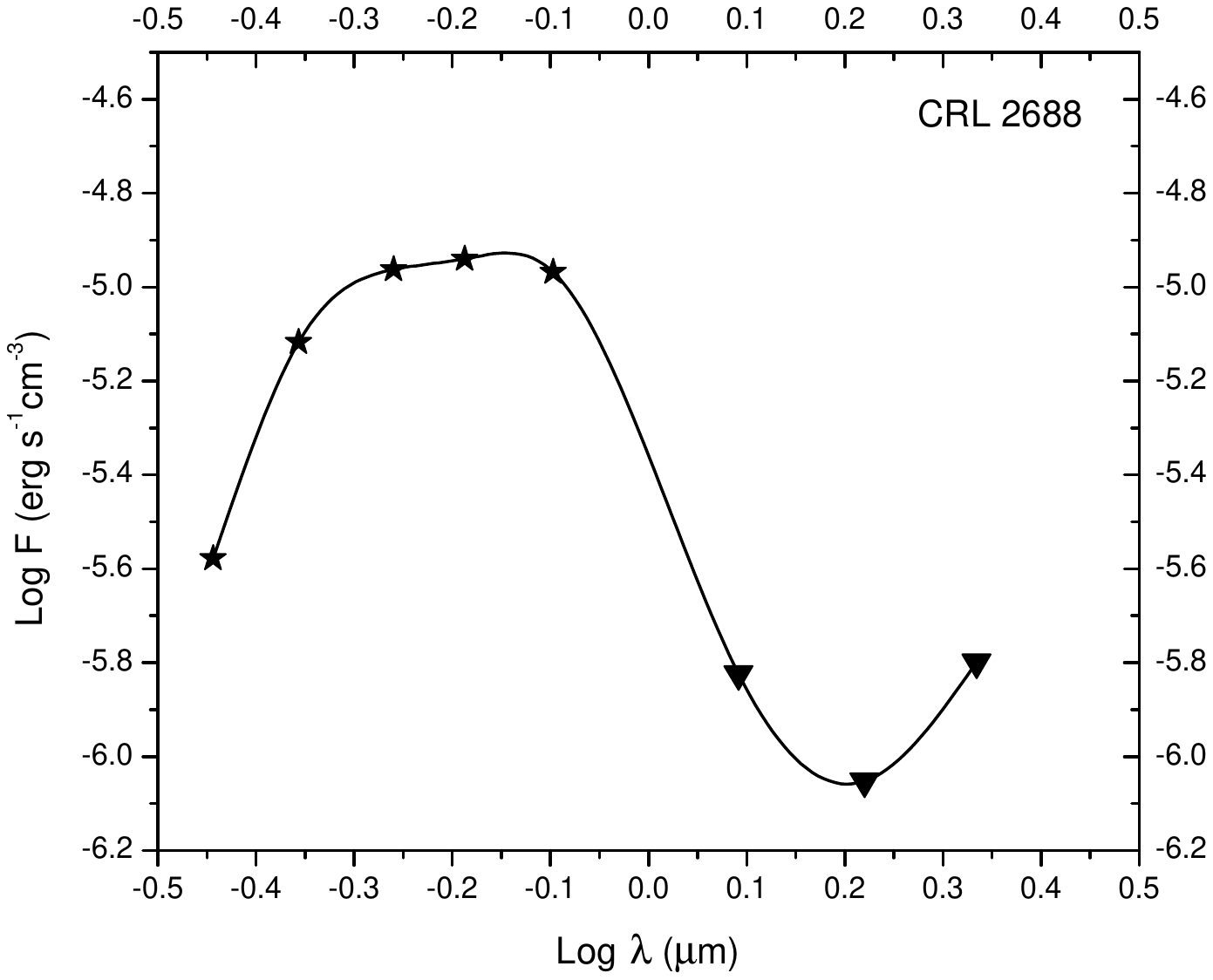}
    \includegraphics[width=0.48\textwidth]{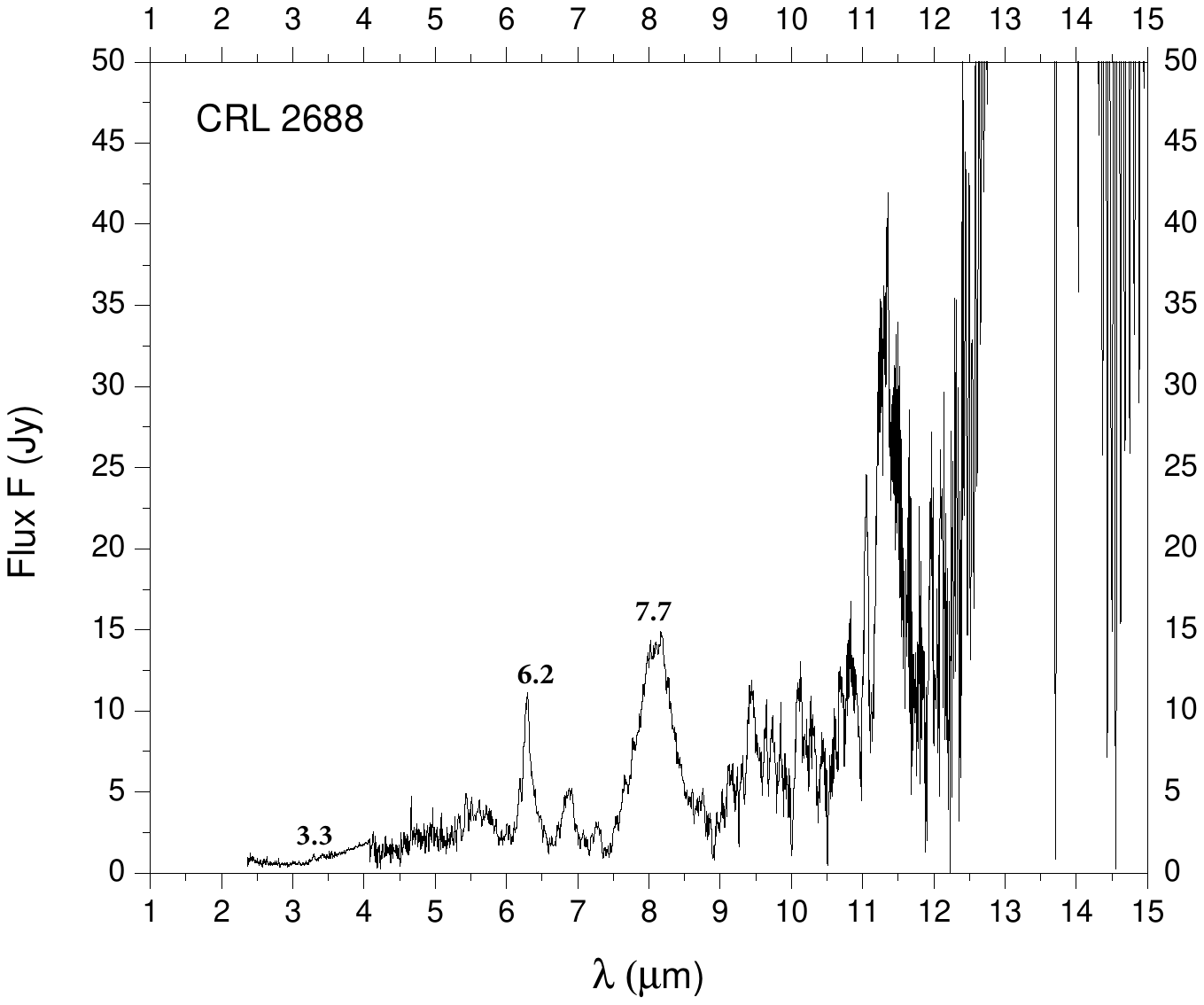}
      \caption{ \textbf{CRL~2688} Upper panels: Near-infrared imaging at various wavelengths. PAH-nbL image presents the region in which PAH is present (circled in black). Lower panels: UBVRIJHK spectral energy distribution (left), UBVRI are denoted with $\star$ while JHK is denoted with $\blacktriangledown$, and background-subtracted mid-IR emission features from ISO SWS (right), major aromatic bands are denoted in spectra in which 7.7$\mu$m and 8.6$\mu$m are blended with each other.}
\label{fig-CRL2688}
\end{figure*}

Optical (UBVRI) images clearly reveal a double-lobed structure resulting from scattering of the central star's radiation by surrounding dust. Near-infrared (JHK) images further emphasize this bipolar morphology; however, the central star remains obscured even at these wavelengths. The northern lobe appears brighter and extends approximately 15~arcsec, whereas the southern lobe is fainter and extends about 7~arcsec (Figure~\ref{fig-CRL2688}). The object shows strong continuum emission in the nbL band (3.59~$\mu$m), and notably, the PAH emission at 3.28~$\mu$m is detected predominantly in the lower lobe of the nebula.

Spectroscopic analysis reveals  AIBs at 3.3, 6.2, and 11.2~$\mu$m, along with a prominent feature near 8.1~$\mu$m. However, emission bands typically found at 7.7 and 8.6~$\mu$m are blended with each other (Figure~\ref{fig-CRL2688}). The environment of CRL~2688 is molecular-rich and relatively shielded from intense radiation fields, facilitating the presence of neutral PAHs, which typically exhibit weak or absent C–C stretching and C–C–H bending modes in the 6--9~$\mu$m region \citep{Pathak2008}. Following the classification scheme by \citet{Diedenhoven2004}, the 3.3~$\mu$m band belongs to class `A', whereas other detected AIBs are classified as type `C'.

The SED of CRL~2688, constructed from UBVRIJHK photometry (Figure~\ref{fig-CRL2688}), clearly demonstrates a pronounced infrared excess indicative of thermal emission from circumstellar dust heated by the embedded star. This SED underscores the significant dust contribution to the nebula’s emission.

Considering the intrinsic colors of an F5Ia star (e.g., Mirfak, $(B-V)_0\approx+0.48$; \citealt{Johnson1966}), we derive substantial reddening toward CRL~2688. Observed colors and derived color excesses are summarized in Table~\ref{tab:combined_colors}, and extinction estimates are summarized in Table~\ref{tab:av_extinction_summary}.

These derived extinctions represent primarily the interstellar component affecting the stellar continuum. The observed large infrared excess ($E(J-K)\sim2.2$~mag) significantly exceeds the reddening expected purely from extinction, indicating dominant thermal emission from circumstellar dust rather than extinction alone. 

Literature estimates of total extinction toward CRL~2688 report significantly higher values due to the nebula’s dense circumstellar environment. Polarimetric studies by \citet{Shawl1976} suggest a total color excess up to $E(B-V)\approx1.7$, corresponding to substantially higher line-of-sight extinction ($A_V\sim5$--6~mag). This discrepancy arises from our detection of only scattered stellar radiation, whereas direct starlight is heavily obscured by the nebula's equatorial dust torus. Thus, our calculated extinction values ($A_V\sim1.5$~mag) effectively represent the foreground interstellar component combined with a small fraction of circumstellar scattering, whereas true line-of-sight extinction is considerably greater.

\subsection{PN M 2-43}
\begin{figure*}[h]
     \centering
     \begin{subfigure}[b]{0.24\textwidth}
         \includegraphics[width=\textwidth]{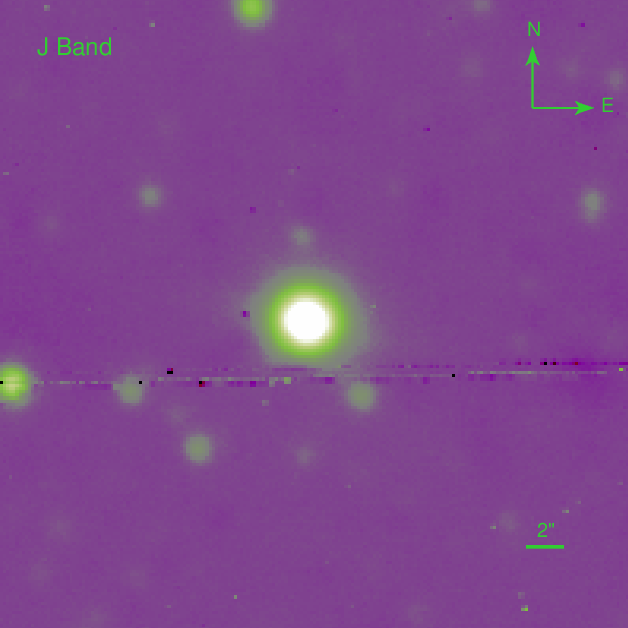}
        
     \end{subfigure}
     \hfill
     \begin{subfigure}[b]{0.24\textwidth}
         \includegraphics[width=\textwidth]{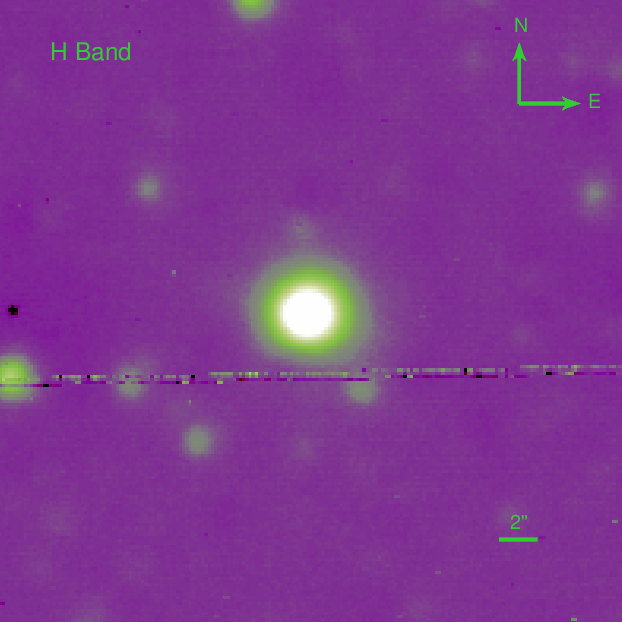}
        
     \end{subfigure}
     \hfill
     \begin{subfigure}[b]{0.24\textwidth}
         \includegraphics[width=\textwidth]{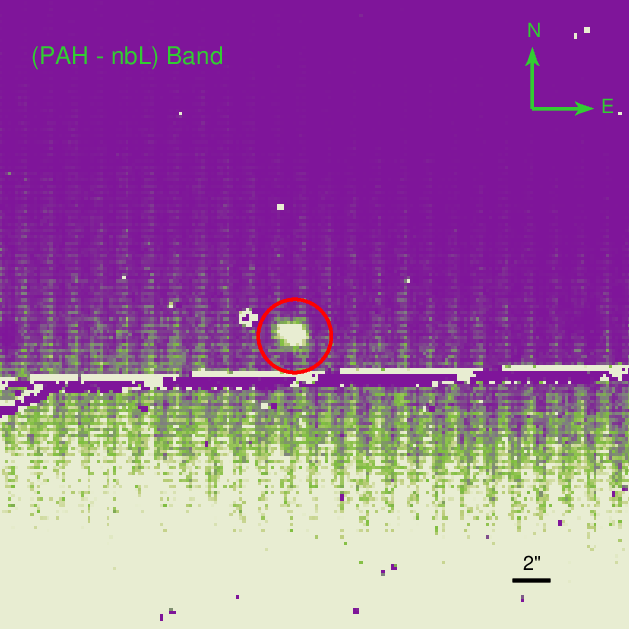}
         
     \end{subfigure}
     \hfill
     \begin{subfigure}[b]{0.24\textwidth}
         \includegraphics[width=\textwidth]{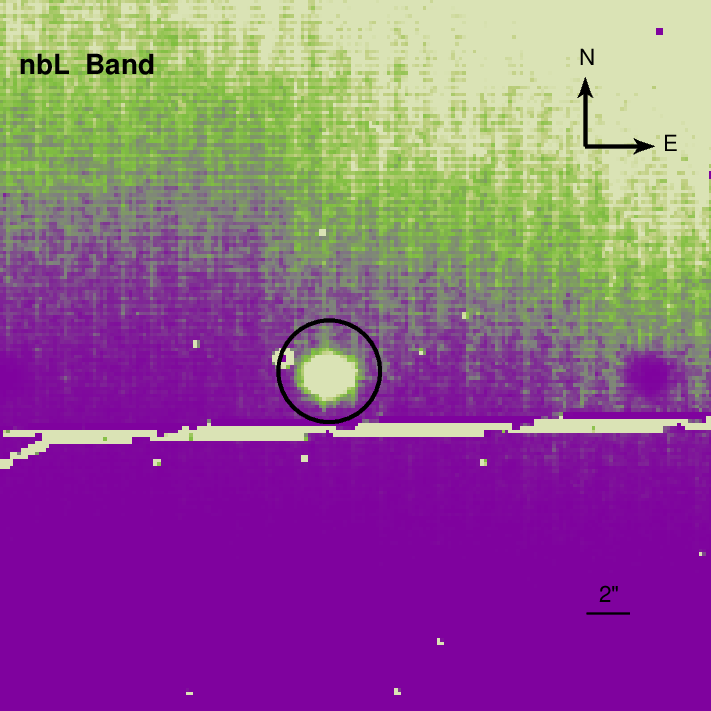}
        
     \end{subfigure}  
    \includegraphics[width=0.48\textwidth]{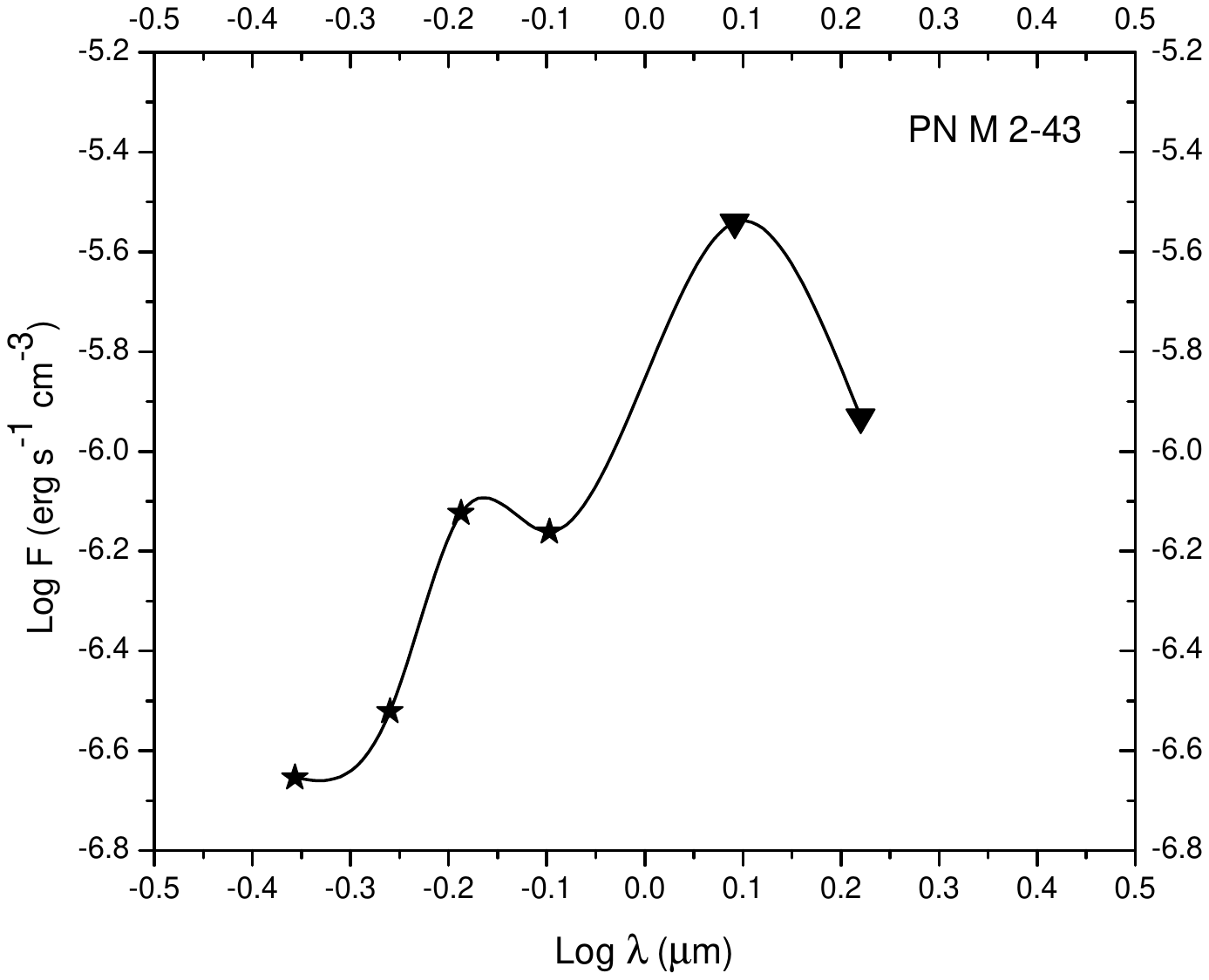}
    \includegraphics[width=0.48\textwidth]{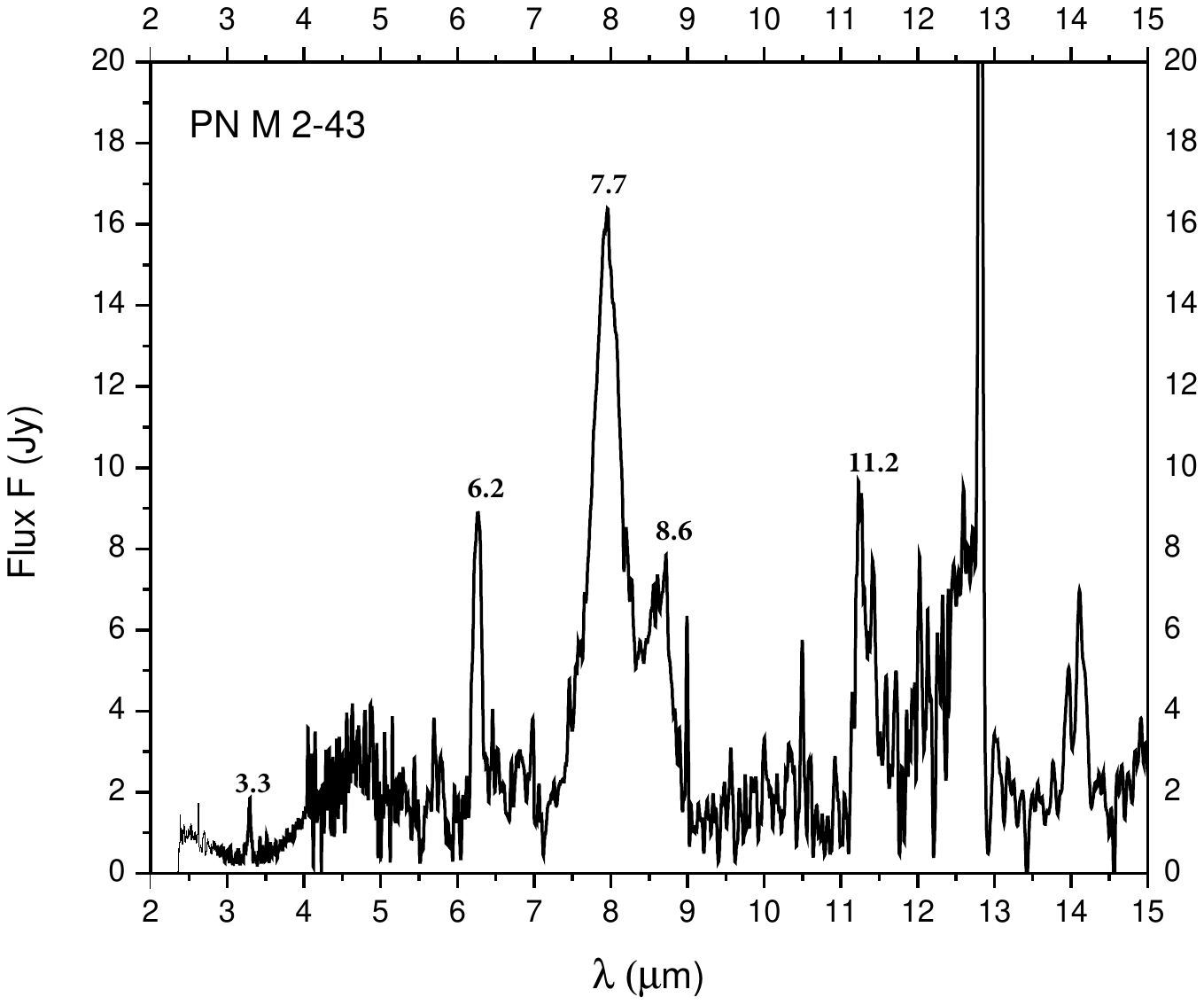}
      \caption{\textbf{PN M~2-43} Upper panels: Near-infrared imaging at various wavelengths. PAH-nbl image presents the clear presence of PAH in the region circled in red. Lower panels: BVRIJH spectral energy distribution (left), BVRI is denoted with $\star$ while JH with $\blacktriangledown$, and background-subtracted mid-IR emission features from ISO SWS (right), major aromatic bands are denoted in spectra. }
      \label{fig-pnm2}
\end{figure*}

PN M~2-43 (also designated as PN~G027.6+04.2) is a compact, bipolar planetary nebula with a central star classified as a late-type Wolf–Rayet star of spectral type [WC7–8] \citep{Leuenhagen1998, Acker2003}. This class of central star is characterized by extremely high temperatures (70,000--80,000~K), intense ultraviolet radiation, and strong stellar winds, contributing to a turbulent circumstellar environment \citep{Grosdidier2001}. High-resolution optical (BVRI) and near-infrared (JH) imaging (Figure~\ref{fig-pnm2}) reveals an approximately spherical central structure with an angular diameter of about 4~arcsec, surrounded by an asymmetrical extended nebular envelope. The nebula exhibits strong internal reddening, with measured magnitudes $V=15.19$, $J=10.09$, $H=9.95$, and color indices $(B-V)=+0.92$, $(J-H)=+0.15$ (see Table~\ref{tab:combined_colors}). 

Infrared emission from circumstellar dust dominates the nebula’s spectrum, significantly hindering direct visibility of the central star even at near-infrared wavelengths. The SED constructed from BVRI and JH photometry shows a prominent peak at the $J$-band (Figure~\ref{fig-pnm2}), reflecting substantial dust emission. Continuum-subtracted imaging in the PAH filter at 3.28~$\mu$m confirms the presence of AIB emission, indicative of PAH molecules excited by intense ultraviolet radiation.

The mid-infrared spectrum (Figure~\ref{fig-pnm2}) exhibits strong AIB features at 3.3, 6.2, 7.7, 8.6, and 11.2~$\mu$m. All observed AIBs closely resemble class `A' profiles, typically associated with young stellar objects and environments characterized by energetic radiation fields and turbulent outflows \citep{Rinehart2002, Grosdidier2001}. Such conditions are common around Wolf-Rayet type planetary nebula nuclei, consistent with the observed properties of PN M~2-43.

Given the intrinsic blue continuum of the central [WC7--8] star (intrinsic $(B-V)_0\approx-0.3$; \citealt{Feinstein1965, Vanderhucht2001}), the observed optical and infrared colors reveal significant reddening. Observed photometric colors and derived color excesses are summarized in Table~\ref{tab:combined_colors}. The optical color index $(B-V)=+0.92$ yields a color excess of $E(B-V)=1.22$~mag, indicating substantial foreground extinction. The extinction estimates are summarized in Table~\ref{tab:av_extinction_summary}.

The observed infrared excess ($E(J-K)\approx2.43$ mag) greatly exceeds the expected reddening solely due to extinction (approximately $0.17 A_V\approx0.65$ mag). This discrepancy clearly indicates that the nebular emission (including Br$\gamma$ and hot dust continuum emission) significantly contributes to the near-infrared flux, particularly in the $K$-band. Therefore, our extinction values represent the foreground interstellar component affecting the stellar continuum, while the nebula’s local emission processes enhance the observed infrared flux significantly.

In comparison to the literature, continuum-based extinction measurements for PN M~2-43 are relatively scarce, as studies typically derive extinction using nebular emission-line ratios. Literature reports highlight PN M~2-43 as among the most highly obscured planetary nebulae, often citing an extinction coefficient $c$(H$\beta$)$\approx1.7$--$2.0$, roughly corresponding to $E(B-V)\approx1.2$--$1.4$ mag \citep{Leuenhagen1998, Acker2003}. These results align well with our derived optical extinction ($A_V\approx3.8$ mag), confirming that PN M~2-43 is significantly extinguished along the line of sight, although its broadband colors are further complicated by strong nebular and dust emission components.

\subsection{NGC 7027}

NGC~7027 is a young, dense planetary nebula and one of the brightest planetary nebulae observable from Earth \citep{Salas2001,kaler2002}. Its central star is extremely hot, with an estimated effective temperature of about $200,000$~K, intrinsically exhibiting a very blue continuum ($(B-V)_0 \approx -0.3$ or bluer). The nebula, located close to the Galactic plane, experiences significant interstellar extinction. Its structure is complex, comprising an elliptical region of ionized gas surrounded by intricate dust structures \citep{Kastner2001}. Optical photometry reveals bright emission across the UBVRI bands, with a measured visual magnitude $V=8.52$ and color index $(B-V)=+0.77$ (Table~\ref{tab:combined_colors}), indicative of substantial reddening along the line of sight.

\begin{figure*}[h]
    \centering
         \begin{subfigure}[b]{0.24\textwidth}
         \centering
         \includegraphics[width=\textwidth]{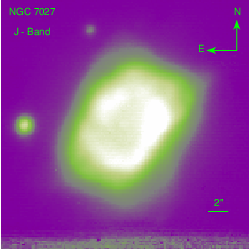}
         \label{7027_J-Band}
     \end{subfigure}
     \hfill
     \begin{subfigure}[b]{0.24\textwidth}
         \centering
         \includegraphics[width=\textwidth]{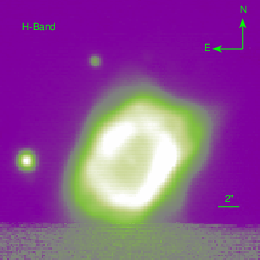}
         \label{ngc_H-Band}
     \end{subfigure}
     \hfill
     \begin{subfigure}[b]{0.24\textwidth}
         \centering
         \includegraphics[width=\textwidth]{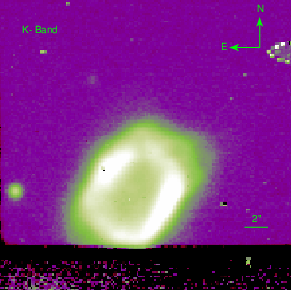}
         \label{7027_PAH-NBL-Band}
     \end{subfigure}
     \hfill
     \begin{subfigure}[b]{0.24\textwidth}
         \includegraphics[width=\textwidth]{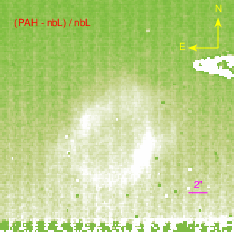}
         \label{7027_pah-Band}
         \end{subfigure}
    \includegraphics[width=0.48\textwidth]{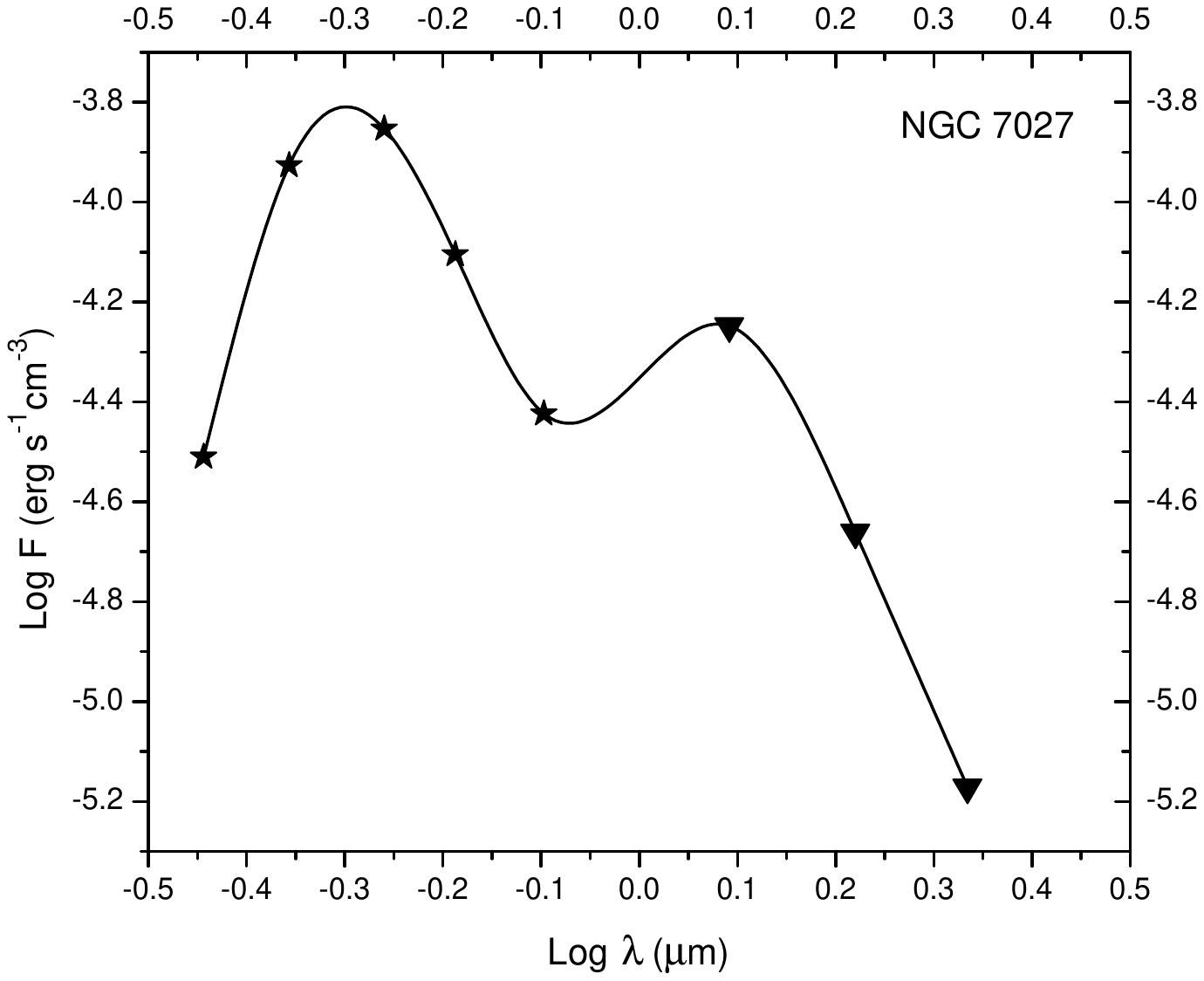}
        \includegraphics[width=0.48\textwidth]{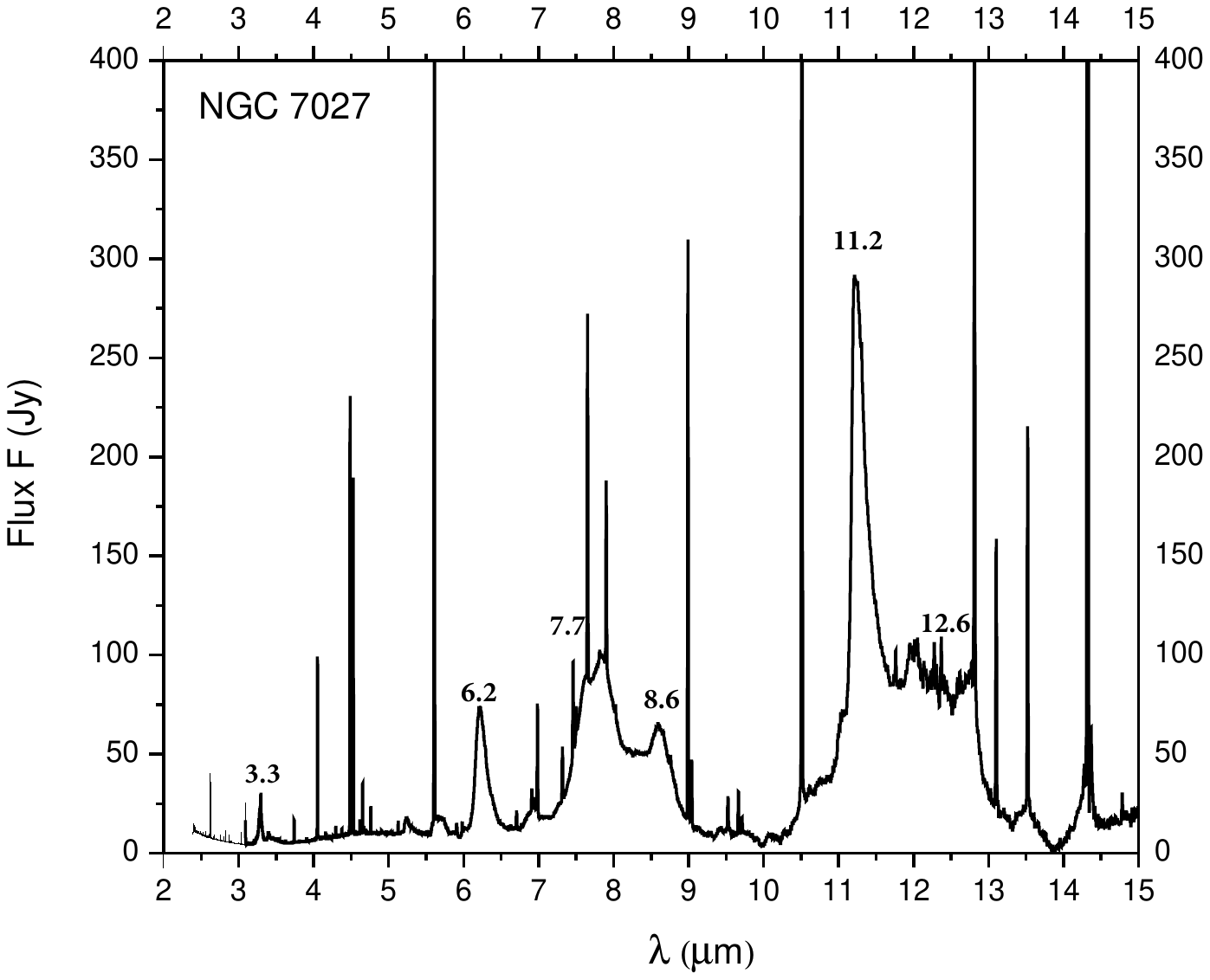}
    \caption{\textbf{NGC~7027} Upper panels: Near-infrared imaging at various wavelengths (taken from \cite{Anand2020}). PAH-nbL image represents the region where PAH is observed in PN. Lower panels: UBVRIJHK spectral energy distribution, UBVRI is represented by $\star$, and JHK is represented by $\blacktriangledown$. Right panel: Background-subtracted mid-infrared emission features from ISO SWS observations, major aromatic bands are represented in the spectra. }
    \label{fig-7027}
\end{figure*}

Near-infrared photometric and narrow-band (nbL and PAH) observations were previously reported by \citet{Anand2020}. The combined UBVRIJHK SED shows that flux densities decrease steadily toward longer wavelengths (Figure~\ref{fig-7027}), suggesting a decreasing contribution from the central star's hot continuum and increasing dominance of nebular emission and circumstellar dust.

Mid-infrared spectroscopy obtained with ISO SWS (Figure~\ref{fig-7027}) exhibits strong AIB emission features at 3.3, 6.2, 8.6, and 11.2~$\mu$m. All observed features in NGC~7027 show relatively high flux compared to other objects in our sample, reflecting a significant abundance and excitation of PAHs within dense, dust-rich regions. Specifically, the AIB features at 6.2~$\mu$m and 7.7~$\mu$m correspond to classes `A' and `B', respectively, typical of regions shielded by dust, promoting active molecular chemistry and strong PAH excitation.

Given the intrinsic blue continuum expected for the hot central star, we derive significant extinction based on the observed optical colors. However, broadband photometry for NGC~7027 is heavily influenced by intense nebular emission lines (e.g., [O\,{\sc iii}], H$\beta$, He\,{\sc ii}), which complicate extinction estimates from continuum colors alone. Notably, colors such as $V-R$ and $R-I$ display unusual negative values due to strong emission lines in the $R$ band. Therefore, the $B-V$ color remains our primary indicator for estimating the continuum extinction.
Observed colors, intrinsic assumptions, and calculated excesses are summarized in Table~\ref{tab:combined_colors}. The extinction estimates are summarized in Table~\ref{tab:av_extinction_summary}.

Calculated values represent the interstellar extinction component affecting the nebula’s stellar continuum. However, extinction measurements derived from nebular hydrogen recombination lines typically yield slightly lower values due to reduced contamination by nebular line emissions. Literature studies based on the Balmer decrement report values around $E(B-V)\approx0.90$ ($A_V\approx2.8$ mag; \citealt{Latter2000, Middlemass1990}). Our slightly higher extinction estimate ($A_V\approx3.3$ mag) likely arises from broadband contamination by emission lines.

Our derived extinction aligns reasonably well with previously published values, typically in the range of $A_V\approx2.8$--$3.0$ mag \citep{Latter2000, Middlemass1990}. Small differences between the literature and our results are attributable to complexities inherent in broadband photometry in emission-rich planetary nebulae. In summary, our analysis supports an extinction value of approximately $A_V=3.3$ mag for NGC~7027, confirming substantial reddening consistent with its Galactic plane location and dense circumstellar environment.

\subsection{BD+30\textdegree~3639}
BD~$+$30$^\circ$~3639 (Campbell’s Star) is a bright, compact planetary nebula characterized by a highly symmetrical, round morphology with relatively low excitation \citep{Balick1987}. Its central star is classified as spectral type [WC9], indicative of a relatively cool Wolf–Rayet–type nucleus with an effective temperature around 30,000--40,000~K \citep{Mendez1991}. Photometric measurements across UBVRI bands reveal that the nebula is exceptionally bright, with visual magnitude $V=9.86$ and unusual negative color indices: $(U-B)=-0.61$ and $(B-V)=-0.28$. These negative colors are indicative of strong nebular emission-line contamination rather than solely intrinsic stellar continuum properties.

\begin{figure*}[h]
    \centering
    \begin{subfigure}[b]{0.24\textwidth}
         \centering
         \includegraphics[width=\textwidth]{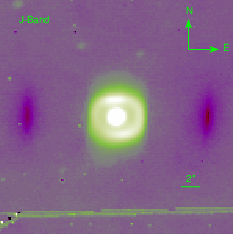}
         \label{BD_J-Band}
     \end{subfigure}
     \hfill
     \begin{subfigure}[b]{0.24\textwidth}
         \centering
         \includegraphics[width=\textwidth]{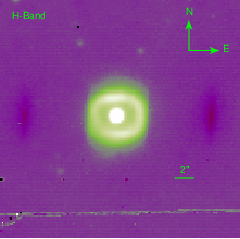}
         \label{BD_H-Band}
     \end{subfigure}
     \hfill
     \begin{subfigure}[b]{0.24\textwidth}
         \centering
         \includegraphics[width=\textwidth]{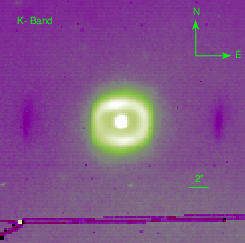}
         \label{BD_PAH-NBL-Band}
     \end{subfigure}
     \hfill
     \begin{subfigure}[b]{0.24\textwidth}
         \includegraphics[width=\textwidth]{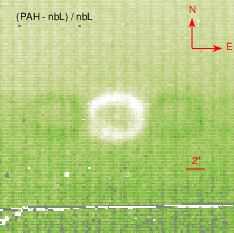}
         \label{BD_pah-Band}
         \end{subfigure}
    \includegraphics[width=0.48\textwidth]{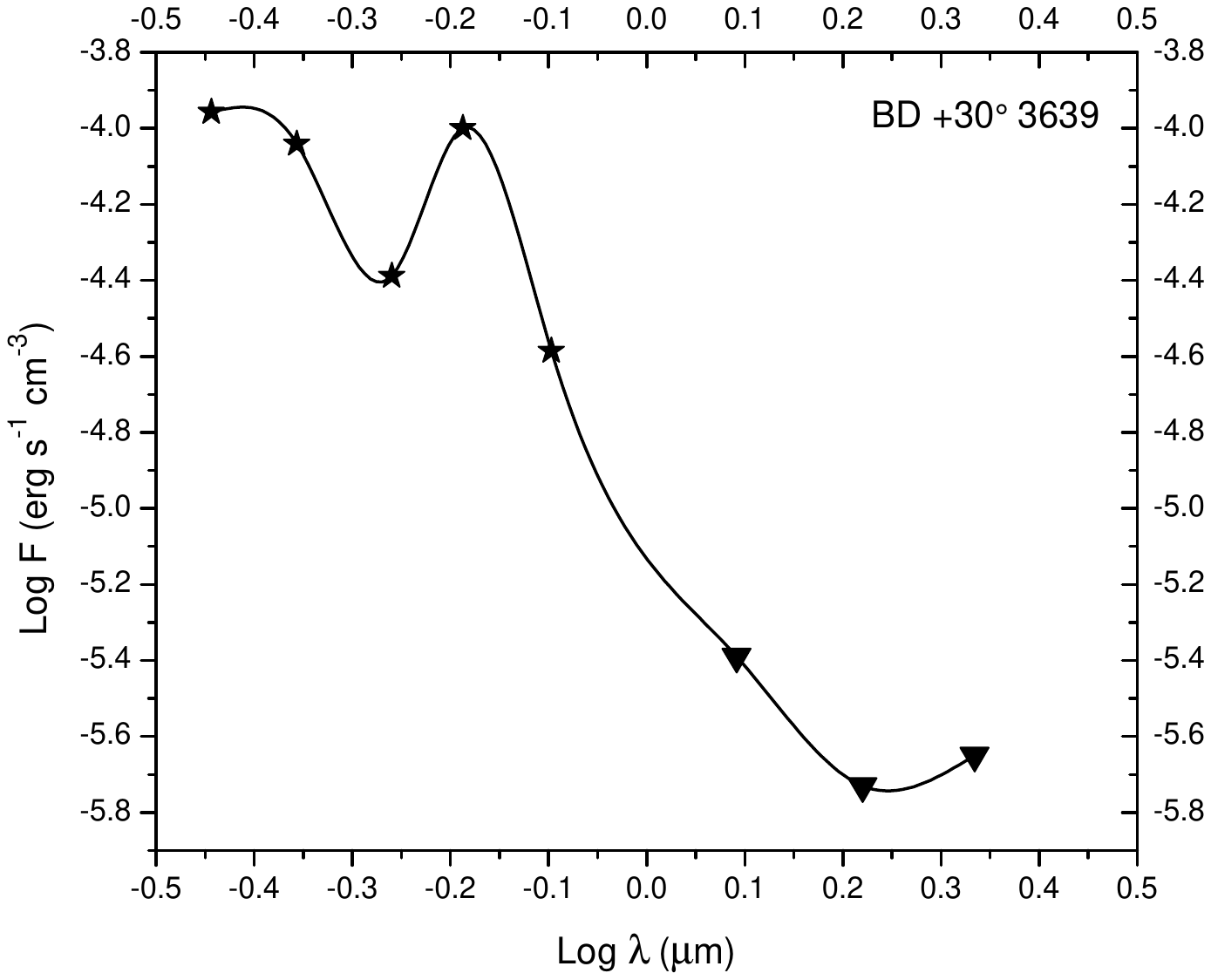}
     \includegraphics[width=0.48\textwidth]{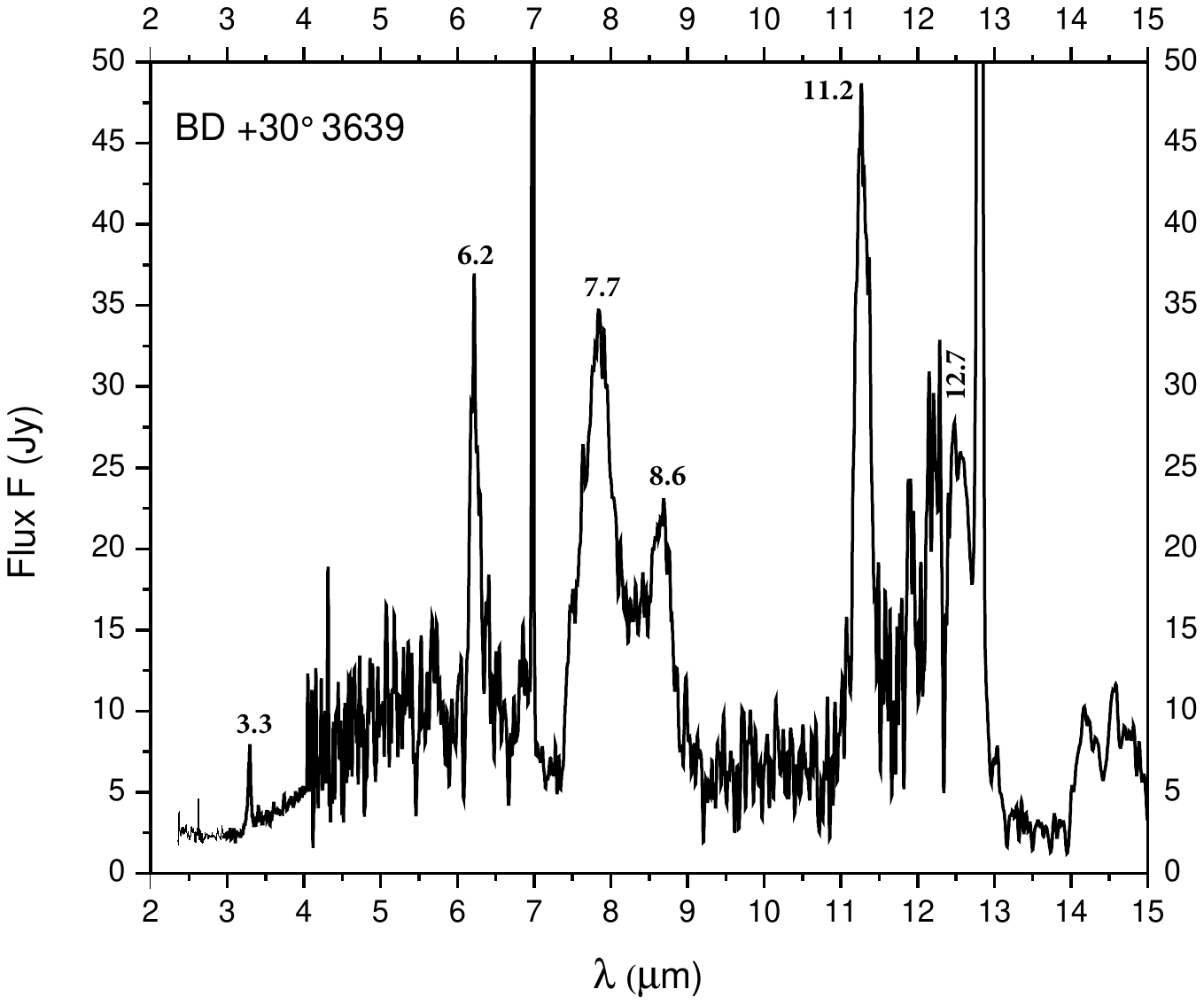}
    \caption{\textbf{BD~$+$30$^\circ$~3639} Upper panels: Near-infrared imaging of BD~$+$30$^\circ$~3639 at various wavelengths (taken from \cite{Anand2020}). PAH-nbL image represents the region where PAH is observed in PN. Lower panels: UBVRIJHK spectral energy distribution (left), UBVRI is represented by $\star$, and JHK is represented by $\blacktriangledown$. Background-subtracted mid-infrared emission features from ISO SWS observations (right), major aromatic bands are indicated in the spectra.}
    \label{fig-BD}
\end{figure*}

Near-infrared observations (JHK), including data from PAH and narrow-band nbL filters, were previously reported by \citet{Anand2020}. The combined UBVRIJHK SED shows a smooth decrease in flux density with increasing wavelength (Figure~\ref{fig-BD}), consistent with substantial emission from the hot central star and surrounding nebula.

Mid-infrared spectroscopy from ISO SWS (Figure~\ref{fig-BD}) displays prominent AIB emission features. BD~$+$30$^\circ$~3639 is historically significant as one of the earliest known planetary nebulae exhibiting strong PAH features, marking it as a benign site of PAH molecule formation. The observed AIB emission at 3.3~$\mu$m corresponds to class `A', while bands at longer wavelengths fall into class `B'. The 11.2~$\mu$m C--H out-of-plane bending feature is notably stronger outside the nebular shell \citep{Matsumoto2008}, indicating partial destruction of PAHs within the nebular shell, likely due to shock interactions.

Given the nebula’s emission-rich nature, broad-band photometry—especially the negative $(B-V)$ index—is significantly influenced by nebular emission lines such as C\,{\sc iii}] and H$\alpha$. Thus, direct extinction estimates based on optical continuum colors are unreliable for BD~$+$30$^\circ$~3639. Instead, we adopt extinction values from more robust literature measurements based on mid-infrared hydrogen recombination lines \citep{Bernard-Salas2003}. 

Observed photometric colors and corresponding intrinsic estimates are summarized in Table~\ref{tab:combined_colors}. The substantial $(J-K)$ color excess strongly suggests significant infrared emission from circumstellar dust.

Adopting the literature-derived optical extinction $E(B-V)=0.34$ from mid-infrared hydrogen lines \citep{Bernard-Salas2003}, we estimate the visual extinction $A_V$, and the extinction estimates are summarized in Table~\ref{tab:av_extinction_summary}.

Our adopted extinction estimate ($A_V\approx1.05$~mag) closely matches previously published measurements. Earlier optical emission-line studies yielded extinction constants corresponding to $E(B-V)\approx0.27$--$0.30$, consistent with our adopted mid-infrared-based value \citep{Bernard-Salas2003}. Thus, our extinction values confirm that BD~$+$30$^\circ$~3639 experiences moderate extinction, consistent with its physical proximity and circumstellar environment.

The results reported on the observations in NIR are among the first phase of observations using the ground-based 3.6m Devasthal optical telescope. The positive detections in the 3.28 $\mu$m filter demonstrate the usefulness of the facility. The 3.28 $\mu$m AIB results from the C-H stretching vibration in PAH molecules. The presence of emission in this band indicates the presence of PAHs in the surrounding medium of the astrophysical object. The C-H out-of-plane motion results in the 11.2 $\mu$m feature, and there is a general correlation with the other C-H vibration; i.e., C-H stretch band \citep{Tielens2008}. The 6.2 $\mu$m AIB is due to the stretching of aromatic C-C bonds in PAHs. The number of C-H bands in the periphery is proportional to the size of the PAHs, which will affect the intensity of the AIBs. In harsh environments, PAHs tend to get more ionized and dehydrogenated. Therefore, the ratio of intensity of the 11.2 and 6.2 $\mu$m bands gives indications of the size and ionization states of possible PAHs in an object. The flux density ratio between 11.2 to 6.2 $\mu$m using the ISO spectral data is tabulated in Table~\ref{tbl-aromatic_bands}.

The flux ratio between the two bands, 11.2/6.2, is $<1$ in AFGL 2132. This points towards a smaller C-H contribution in AFGL 2132. Further, the fact that C-C stretch modes are enhanced in intensity in cations implies that there are more ionized and/or dehydrogenated PAHs in AFGL 2132. The 3.28 $\mu$m band is absent or is weak and undetectable in the NIR observations of this object (Figure~\ref{fig-afgl2132}). 

The other four objects have $11.2/6.2 >1$, with the largest ratio 3.946 in NGC 7027. NGC 7027 provides strong background excitation \citep{Latter2000}, enhancing all AIBs. This is also considered to be a site for PAH formation and contains large PAHs. This ratio is 2.833 in CRL 2688, which provides a benign environment and is also a possible PAH formation and growth region. Large and newly formed PAHs are also responsible for their mostly class `C' AIB features \citep{Rastogi2013}. In PN~M~2-43, this ratio is nearly one, implying an even distribution of neutral and ionized PAHs. All AIBs in this object belong to class `A'.

\section{Discussion}
\subsection{Extinction and Circumstellar Nebula Characteristics}
 All five nebulae show significant foreground extinction, with visual extinctions ranging from relatively moderate to very high. We derived $A_V \approx 1.0$–1.5 mag for BD ${+}$30${^\circ}$ 3639 (consistent with its known modest extinction) and $A_V \approx 3$–4 mag for the young PNe M~2-43 and NGC~7027, up to $A_V \sim 6$ mag for AFGL~2132 (the most heavily obscured object in our sample). These values reflect the varying column densities of dust along the line of sight. In the case of AFGL~2132, the extraordinarily high extinction confirms that we are viewing this nebula through a dense circumstellar dust layer, likely an edge-on toroidal envelope, which is in line with its nickname “Red Square” for its deeply reddened appearance. For the other nebulae, the extinctions we determined are in good agreement with or somewhat higher than literature values, suggesting that we have captured both interstellar and circumstellar contributions. We also found that simple application of a standard extinction law could not fully account for the observed infrared colors in some objects, notably PN~M~2-43 and NGC~7027, which showed excess emission in the $K$ (2.2~$\mu$m) and $L$ (3–4~$\mu$m) bands beyond what would be expected from extinction alone. This indicates that a significant fraction of their near-IR flux arises from local nebular emission (e.g., hydrogen recombination lines like Br$\gamma$ and warm dust continuum) rather than just the attenuated starlight \citep{Cox2016}. In other words, the dusty nebula itself contributes to the observed infrared brightness. After accounting for this effect, the derived extinctions predominantly represent the foreground (interstellar) component affecting the stellar continuum, while the intrinsic nebular dust emission is recognized as a separate contributor.

\subsection{Prominent PAH Emission in All Objects}
 The mid-infrared spectra of all five nebulae exhibit the characteristic PAH/AIB features, confirming active carbon chemistry in their circumstellar shells. We detect strong emission bands at 6.2, 7.7, 8.6, and 11.2$\mu$m in each object’s ISO spectrum, and the 3.3 $\mu$m aromatic C–H stretch feature is present in four of the five objects. Notably, AFGL~2132 is the only object in our sample where the 3.3 $\mu$m PAH feature is absent or extremely weak. Given that 3.3$\mu$m emission typically traces smaller PAH molecules (which have more C–H per C atom and tend to emit strongly at 3.3$\mu$m when excited), its non-detection in AFGL~2132 suggests an anomalous PAH population in that nebula. One possibility is that the PAH molecules in AFGL~2132 are predominantly larger and perhaps more graphitic, lacking the smaller aromatics that produce the 3.3$\mu$m band. Another possibility is that the intense local conditions (e.g., extreme extinction or perhaps a peculiar radiation field) suppress the excitation of the 3.3$\mu$m band. By contrast, the other nebulae all show a clear 3.3$\mu$m feature, indicating the presence of a normal complement of small-to-medium PAHs. This dichotomy makes AFGL~2132 stand out as a case of “altered PAH chemistry,” qualitatively consistent with its unique status and unclear evolutionary state. Apart from the 3.3$\mu$m line, all the nebulae display the full suite of mid-IR PAH features, reinforcing that carbonaceous molecules form and persist in these environments even under the influence of strong UV fields.

\subsection{Variations in PAH Profiles – Neutral vs. Ionized PAHs}
 A key result of our study is that, although each object shows the same set of PAH bands, the relative intensities of those bands differ systematically among the nebulae. We quantified the PAH band ratios – for example, the flux ratio of the 11.2$\mu$m (C–H out-of-plane bending) to the 6.2$\mu$m (C–C stretching) feature, F(11.2)/F(6.2). This ratio serves as a diagnostic of PAH ionization: neutral PAH molecules tend to emit relatively strongly at 11.2$\mu$m, whereas ionized PAHs enhance the 6–8$\mu$m bands. We found that F(11.2)/F(6.2) ranges from $\sim$0.8 in AFGL~2132 up to $\sim$3.9 in NGC~7027 (with intermediate values for the others). In NGC~7027 and PN~M~2-43, the 6.2 and 7.7$\mu$m bands are particularly prominent, and the 11.2 $\mu$m feature, while present, is comparatively weaker (yielding low 11.2/6.2 ratios). This implies that a large fraction of the PAHs in these two nebulae are ionized, not surprising given their central stars are very hot ([WC7]-[WC8] type in the case of NGC~7027) and produce intense UV radiation. The hard radiation field and possibly fast shocks in these nebulae likely lead to a PAH population that is smaller and more charged, consistent with Type A PAH spectra in the classification scheme (common in high-excitation sources). By contrast, CRL~2688 and BD ${+}$30${^\circ}$ 3639 show a different PAH profile: the 11.2 $\mu$m band (and the 12.7$\mu$m PAH band) are relatively strong compared to the 6–8$\mu$m complex, indicating a dominance of neutral or only mildly ionized PAH molecules in these objects. CRL~2688, being a post-AGB object, still has a much cooler central star (not yet a strong UV source), which naturally leads to mainly neutral PAHs surviving in its outflows. BD ${+}$30${^\circ}$ 3639, though it does have a hot central [WC9] star, is known to have a dense, molecular-rich envelope and a relatively lower-ionization nebula; our results support previous findings that its mid-IR spectrum is dominated by emission from larger, neutral PAHs \citep{Anand2020, Anand2023}. In summary, we observe a clear pattern: objects with softer or shielded radiation fields (CRL~2688, BD ${+}$30${^\circ}$ 3639) exhibit PAH spectra characteristic of neutral PAH populations (strong 11.2$\mu$m), whereas those with harder, more intense UV fields (NGC~7027, M~2-43) exhibit PAH spectra skewed towards ionized PAHs (strong 6–8$\mu$m bands). These differences are in line with theoretical expectations and have been noted qualitatively in other studies of PAH-rich PNe (e.g., comparison of PAH emission in various PNe by \citep{Peeters2002}). Our work provides quantitative band ratio measurements that substantiate this paradigm.

\subsection{Correlating Optical Properties with PAH Emission}
One of the motivations of this study was to search for correlations between the nebulae’s optical properties (such as extinction or excitation class) and their PAH feature characteristics, as any such correlations could help connect the macroscopic nebular evolution to the microscopic dust chemistry. We are unable to find a one-to-one correlation between the amount of extinction $A_V$ and the overall strength of PAH emission in the mid-IR – even the heavily obscured AFGL~2132 shows strong mid-IR dust emission (albeit with an unusual PAH spectrum), while the moderately reddened PN~M~2-43 has PAH features as intense as those of the more obscured NGC~7027. This suggests that, within this sample, the quantity of dust (extinction) does not directly govern the PAH spectral output. However, when considering the quality of the PAH spectrum (ionization indicators as discussed above), there is a loose correspondence with nebular evolutionary status and central star type. CRL~2688, which is a very young nebula (central star not yet hot enough to ionize helium), shows PAH features akin to those in proto-planetary nebulae and reflection nebulae (neutral-rich spectra). On the other hand, the young PNe with Wolf-Rayet central stars (NGC~7027, BD ${+}$30${^\circ}$ 3639) show evidence of PAH processing (ionization and perhaps fragmentation). We also note that all objects in our sample are known to exhibit AIB “Class A” or “B” profiles rather than Class C (in the Peeters classification); this is consistent with them being relatively high-excitation sources – even CRL~2688, though a pre-PN, has a central star that has produced an observable UV-pumped PAH spectrum. Another interesting observational result is that the 11.2~$\mu$m PAH flux relative to the thermal continuum appears enhanced in the cooler objects. This might indicate that in cooler environments, PAH emission contributes a larger fraction of the IR output (since less of the dust is at very high temperatures), whereas in extremely hot environments, some fraction of small PAHs might be destroyed or fully ionized (shifting emission to longer-wavelength plateau features or continuum). Overall, the relationship between optical nebular properties and PAH features is complex – no single optical indicator (e.g., $T_\mathrm{eff}$ of the central star or $A_V$) alone predicts the PAH spectrum. Instead, a combination of factors – the central star temperature (hence available UV), the nebular morphology (which determines how much UV reaches the molecules), and the dust abundance and prior history – collectively influence the PAH chemistry. This complexity highlights the need for detailed, object-specific models and suggests caution in generalizing PAH diagnostics across different nebulae.

\subsection{Implications for Dust and PAH Evolution}
Our findings underscore the diverse and complex dust environments present in carbon-rich nebulae. Even among objects at ostensibly similar evolutionary stages (young PNe, a few thousand years old), we see significant variations in their dust extinction and PAH emission properties. This diversity likely reflects differences in initial progenitor mass, mass-loss history, and binarity, which result in different circumstellar densities and geometries at the onset of the PN phase. From a theoretical standpoint, these results provide valuable constraints. For example, any successful model of PAH formation in PNe must accommodate the survival of PAH molecules in very harsh UV conditions (as in NGC~7027, which still shows abundant PAH emission despite a $T_\mathrm{eff}\sim200$ kK central star) while also explaining the rapid emergence of PAHs in the post-AGB stage (as evidenced by CRL~2688). The clear presence of predominantly neutral PAHs in some objects versus highly ionized PAHs in others can be used to calibrate models of PAH ionization in nebulae – essentially serving as empirical tests of how PAHs respond to different radiation field strengths \citep{Anand2020}. Moreover, the measured band ratios (like 11.2/6.2) in our study can be directly compared to photochemical models of PAH mixtures to infer typical PAH sizes in each nebula. For instance, the very high 11.2/6.2 ratio in NGC~7027 $(>3)$ hints at either an abundance of large PAHs (which remain neutral) in shielded clumps within that nebula, or conversely, an overall PAH population that is partially destroyed except for resilient large grains that then emit strongly at 11.2 $\mu$m \citep{Cox2016}. On the other hand, the moderate ratio in M~2-43 (~1) suggests a balance of ionized and neutral PAHs in that object’s knots, consistent with a somewhat softer UV field or shorter PAH survival times.

\section{Conclusion}
Our study shows that all of the targeted nebulae have a lot of foreground visual extinction. For example, BD~${+}30{^\circ}$~3639 has an $A_V$ of about 1.0 mag, while AFGL~2132, which is heavily obscured, has an $A_V$ of about 6 mag. This nebula looks like it is being viewed through a dense, edge-on circumstellar dust torus. We found near-infrared excess emission for some objects, namely PN M 2-43 and NGC 7027. This means that local nebular processes like hydrogen recombination lines and warm dust continuum are adding a lot to the light, in addition to the starlight that has been weakened.

All five nebulae showed strong mid-infrared PAH emission characteristics (6.2, 7.7, 8.6, and 11.2 $\mu$m), which shows that the carbon chemistry in their circumstellar shells is still going on and is quite active. The 3.3 $\mu$m PAH characteristic stood out as a major difference: its lack or extreme weakness in AFGL~2132 shows that this nebula has an atypical PAH population, which may be made up of larger, more graphitic PAHs or may be experiencing strange excitation circumstances. The other four objects, on the other hand, do show this trait, which means they have a more usual mix of tiny to medium PAHs.

It was also important that we saw regular changes in the PAH spectrum profiles, which showed that the ionization states were different. The F(11.2)/F(6.2) band ratio was a very important diagnostic:
\begin{itemize}
    \item \textbf{CRL~2688 and BD~${+}$30${^\circ}$~3639} displayed spectra characteristic of predominantly neutral PAHs, consistent with their softer or more shielded radiation fields.
    \item \textbf{NGC~7027 and PN~M~2-43} showed spectra dominated by ionized PAHs, attributable to the intense ultraviolet radiation from their hot central stars. AFGL~2132 exhibited a low F(11.2)/F(6.2) ratio, alongside its unique 3.3 $\mu$m characteristic.
\end{itemize}

There wasn't a direct one-to-one link between the total extinction $A_V$ and the overall strength of PAH emission, but the nature of the PAH spectrum (ionization balance) does show a clear, though complicated, link with the evolutionary status of the nebula and the properties of the central star. This shows that the observed PAH chemistry is controlled by a number of factors, such as the temperature of the central star, the shape of the nebula, and the amount of dust.

In short, this study shows that there is a lot of variety in dust environments and PAH processing, even in nebulae that seem to be at the same stage of evolution. The quantitative band ratios and characterized PAH populations give theoretical models strong real-world standards. These results are very important for improving our understanding of how PAHs develop, survive, ionize, and change size in response to different astrophysical conditions. They are also important for making PAH characteristics more accurate as diagnostic tools in carbon-rich circumstellar environments.

\section*{Acknowledgements}
We are grateful to the observing and technical staff of the 1.3-m Devasthal Fast Optical Telescope (DFOT) and the 3.6-m Devasthal Optical Telescope (DOT) at ARIES, Nainital, for their assistance during the observations. We also acknowledge the use of archival data from the ISO/SWS database, and we thank the developers of the associated calibration and analysis tools. 

A.K.S. acknowledges financial support through the National Fellowship for Other Backward Classes (NFOBC), awarded by the Ministry of Social Justice and Empowerment, Government of India.

\bibliographystyle{elsarticle-harv} 
\bibliography{the_bibtex}

\end{document}